\documentclass[acmsmall]{acmart}

\usepackage{comment}
\usepackage{listings}
\usepackage{xcolor}
\usepackage{balance}
\usepackage[most,skins]{tcolorbox}
\usepackage{float}
\usepackage{graphicx}
\usepackage{placeins}
\usepackage{enumitem}
\usepackage{multirow}
\usepackage{bm}
\usepackage{url}
\usepackage{hyperref}
\usepackage{xspace}
\usepackage{algorithm}
\usepackage{algorithmic}
\usepackage{booktabs}
\usepackage{fancybox}
\usepackage{subcaption}
\usepackage{colortbl}
\usepackage{rotating}
\usepackage{makecell}

\setcopyright{cc}
\setcctype{by}
\acmJournal{PACMSE}
\acmYear{2026} \acmVolume{3} \acmNumber{ISSTA} \acmArticle{ISSTA011}
\acmMonth{10} \acmDOI{10.1145/3832102}
\acmSubmissionID{issta26main-p101-p}
\received{2026-01-30}
\received[accepted]{2026-06-25}

\newcommand{\toolname}{\textsc{Delta}\xspace}

\keywords{Deep Reinforcement Learning, Differential Testing, Fuzz Testing}

\begin{document}

\begin{abstract}
    Deep Reinforcement Learning (DRL) has achieved significant success in complex decision-making problems. As DRL systems are increasingly deployed in real-world applications, ensuring their quality and reliability is paramount. 
    Current works primarily focus on detecting safety-critical failures, often neglecting policy optimality, which can lead to reduced efficiency, user distrust, and economic losses. This oversight, compounded by the inherent ``testing oracle problem'' for optimality, leaves a significant gap in comprehensively evaluating DRL systems. To address this gap, we propose \toolname (Differential Testing for DRL Agents), a novel and comprehensive framework that automatically identifies both safety-critical and optimality bugs in DRL agents. \toolname employs a two-phase approach: (1) Safety Testing, where the Agent Under Test (AUT) is evaluated for catastrophic failures while collecting data from its decision-making policy, and (2) Optimality Testing, where this collected data from the prior phase is used to train a \emph{challenger} agent via Offline Reinforcement Learning. Differential testing is then performed by comparing the challenger agent against the AUT; instances where the challenger achieves higher cumulative rewards indicate optimality issues in the AUT. We demonstrate \toolname's effectiveness across five environments. We investigate the effectiveness of three offline RL algorithms (BC, BCQ, and CQL) in generating challenger agents. Experimental results demonstrate that safety testing datasets are valuable for training competent DRL agents. Challenger agents trained with BCQ proved most effective for identifying optimality issues within the framework of \toolname. Across the five environments, \toolname uncovered an average of 2,518 optimality issues, outperforming the baseline methods by 50.2\%.
\end{abstract}

\title{Learning from the Test: Self-Referential Differential Testing for Deep RL Agents}

\author{Junda He}
\orcid{0000-0003-3370-8585}
\email{jundahe.2022@phdcs.smu.edu.sg}
\affiliation{%
  \institution{Singapore Management University}
  \city{Singapore}
  \country{Singapore}
}

\author{Jieke Shi}
\orcid{0000-0002-0799-5018}
\email{jiekeshi@smu.edu.sg}
\affiliation{%
  \institution{Singapore Management University}
  \city{Singapore}
  \country{Singapore}
}

\author{Zhou Yang}
\orcid{0000-0001-5938-1918}
\email{zy25@ualberta.ca}
\affiliation{%
  \institution{University of Alberta}
  \city{Edmonton}
  \country{Canada}
}
\affiliation{%
  \institution{Canada CIFAR AI Chair, Amii}
  \city{Edmonton}
  \country{Canada}
}

\author{Mingfei Cheng}
\orcid{0000-0002-8982-1483}
\email{mfcheng.2022@phdcs.smu.edu.sg}
\affiliation{%
  \institution{Singapore Management University}
  \city{Singapore}
  \country{Singapore}
}

\author{David Lo}
\orcid{0000-0002-4367-7201}
\email{davidlo@smu.edu.sg}
\affiliation{%
  \institution{Singapore Management University}
  \city{Singapore}
  \country{Singapore}
}

\ccsdesc[500]{Computing methodologies~Markov decision processes}
\ccsdesc[500]{Software and its engineering~Search-based software engineering}

\maketitle

\section{Introduction}

In recent years, Deep Reinforcement Learning (DRL) \cite{HasseltGS16, MnihKSGAWR13, arulkumaran2017deep} has gained significant attention for its ability to tackle complex decision-making problems. By interacting with unknown environments through trial-and-error, DRL agents learn a decision-making policy to maximize cumulative reward over a sequence of actions. DRL has demonstrated remarkable success in various domains such as healthcare~\cite{yu2021reinforcement,coronato2020reinforcement}, autonomous driving~\cite{kiran2021deep,sallab2017deep,shalev2016safe}, and robotic manipulation~\cite{ibarz2021train,kober2013reinforcement,nguyen2019review}, outperforming traditional control methods by a substantial margin~\cite{matsuo2022deep,fuchs2021super}.

As DRL systems, and AI-based software systems more broadly~\cite{he2025mas}, are increasingly deployed in real-world applications~\cite{talpaert2019exploring, nguyen2020deep}, ensuring their quality and reliability becomes paramount~\cite{yang2022natural}. Recent testing methodologies~\cite{pang2022mdpfuzz, he2024curiosity, li2023generative, shi2025synthify} for DRL systems primarily focus on detecting catastrophic failures (safety-critical issues), e.g., a DRL-controlled robot colliding with obstacles. We refer to these testing methods as \emph{safety testing}. Despite this progress, current safety testing methods largely overlook a critical dimension: the decision optimality of DRL systems. Evaluating policy optimality is as crucial as ensuring safety. For instance, a robotic arm performs a task with unnecessary energy consumption. Such non-optimal DRL decisions can incur higher economic costs, reduce operational efficiency, and decrease user satisfaction. More critically, unresolved optimality issues can increase the risk of escalating into safety-critical hazards; for example, a delivery robot selecting congested routes is more likely to face traffic accidents or battery depletion. True system reliability, therefore, demands both safety and optimality. To bridge the critical gap between safety and optimality evaluation, this paper presents a unified framework for testing DRL agents on detecting both \emph{safety-critical failures} and \emph{optimality issues}.

However, a fundamental challenge in testing the optimality of DRL policies is the absence of a clear testing oracle. DRL is predominantly applied in complex and dynamic environments. In these environments, the optimal solution that achieves the theoretical maximum cumulative reward is either unknown or analytically intractable. Without a ground-truth optimal solution to serve as a benchmark, it becomes difficult to assess the optimality of an agent’s decisions quantitatively. This is akin to verifying the output of a software program without knowing the correct output. 
While manual assessment by domain experts is an option, it is prohibitively time-consuming and expertise-intensive, rendering it infeasible to scale for the vast and diverse scenarios DRL agents encounter. Automated, learned surrogates are therefore increasingly explored as scalable evaluators when a ground-truth oracle is unavailable, e.g., for judging software artifacts~\cite{he2025judge} and assessing the correctness of automatically generated patches~\cite{zhou2024patchcorrect}. The impracticality of manual assessment underscores the critical challenge of addressing the oracle problem and developing automated testing methods for DRL policy optimality.

To overcome this, differential testing~\cite{mckeeman1998differential,evans2007differential} emerges as a promising solution. This widely used software testing technique compares the outputs of multiple implementations of the software, thereby eliminating the need to know the correct output. 
Similarly, we adapt this principle to DRL. For a given DRL agent to be tested---we refer to this agent as the agent under test (AUT)---by comparing the AUT’s cumulative reward against other DRL instances, we can reveal the AUT’s optimality issues if it is outperformed in cumulative reward. However, this approach introduces a new challenge: generating suitable DRL instances for comparison.
We can repeat the training process several times to obtain multiple DRL agents for differential testing; however, developers are typically only interested in testing high-performing DRL agents.
Training such policies requires extensive interaction with the environment, consuming considerable computational resources and substantial time~\cite{dulac2021challenges, ding2020challenges}. For large-scale systems, restarting this process can take weeks or even months, requiring the processing of billions of data points~\cite{dulac2021challenges}.

To address this limitation, we introduce a more efficient approach founded on a core insight: the detailed records of an AUT's decision-making process generated during safety testing are a vibrant and underutilized resource for training highly competent DRL agents. The reasons are:
(1)~existing testing methodologies~\cite{he2024curiosity,pang2022mdpfuzz,li2023generative} employ various mechanisms to generate a broad spectrum of scenarios for evaluating the AUT and triggering diverse agent behaviors. These scenarios often include edge cases that are rarely encountered during standard training, deliberately pushing the AUT into less explored regions.
(2)~Developers are typically only interested in testing high-performing DRL agents. Consequently, the AUT is a well-trained agent. Its behaviors, therefore, constitute a high-quality dataset of successful decisions. Leveraging this data provides a new policy with a strong starting point, jump-starts its training process, and bypasses the need for extensive, and often inefficient, initial exploration from scratch.
(3)~Finally, the resulting dataset offers a mixture of learning signals, which is crucial for robust training. It contains not only the AUT's successful trajectories as strong positive examples but also its documented failures as critical negative examples. This mixture provides clear guidance, teaching a new agent what behaviors to replicate and which to avoid.

As a result, we train a new agent using the AUT's own interaction data captured during safety testing. We name this new agent the \emph{challenger} agent. Because it is trained solely on the AUT’s own history, the challenger agent then serves as a potent self-referential testing oracle for suboptimality detection.
This approach pragmatically relaxes the oracle problem. Acknowledging that true optimal policies are often unknown, if the challenger agent finds a superior solution, it exposes a concrete instance where the AUT's policy can be improved. As a result, this method effectively enables automated optimality testing without requiring a predefined perfect solution.

Notably, this approach yields value regardless of the testing outcome. A superior challenger identifies optimality issues and guides policy improvement. Conversely, a challenger that fails to outperform the AUT validates that the AUT is already near-optimal---without ever knowing what the true optimum is.

In this paper, we introduce \toolname (\textbf{D}iff\textbf{e}rentia\textbf{l} \textbf{T}esting for DRL \textbf{A}gents), a comprehensive framework designed to automatically test DRL agents. \toolname seamlessly integrates established safety testing methods, such as CureFuzz~\cite{he2024curiosity}, MDPFuzz~\cite{pang2022mdpfuzz}, and GMT~\cite{li2023generative}, with our proposed differential testing method, enabling holistic evaluation of agent quality in terms of both catastrophic failures and suboptimal behavior. \toolname consists of two main phases: \textbf{(1) Safety Testing:} The AUT is evaluated across a diverse set of scenarios to identify safety-critical issues. We adopt the state-of-the-art technique CureFuzz~\cite{he2024curiosity} for this phase. Simultaneously, \toolname records the AUT's interaction trajectories, creating a rich dataset that underpins subsequent optimality analysis. \textbf{(2) Optimality Testing:} Using the collected trajectories, \toolname applies Offline Reinforcement Learning~\cite{levine2020offline, kostrikov2021offline} to train a \emph{challenger} agent. We then perform differential testing by comparing the challenger's performance against the original AUT. Cases where the challenger achieves higher cumulative rewards are flagged as optimality issues in the AUT.

To assess \toolname's effectiveness, we conducted evaluations across five diverse environments: three classical control tasks (CartPole, MountainCar, Acrobot) and two MuJoCo locomotion tasks (Hopper, Walker2D). We also evaluated the performance of three offline RL algorithms (BC~\cite{torabi2018behavioral}, BCQ~\cite{fujimoto2019off}, and CQL~\cite{kumar2020conservative}) in generating challenger agents. Experimental results demonstrate that safety testing datasets are valuable for training competent DRL agents. Challenger agents trained with BCQ proved most effective for identifying optimality issues within \toolname. Leveraging the BCQ-trained challenger agents, \toolname uncovered an average of 2,518 optimality issues per environment. \toolname significantly outperforms the
baseline methods and identifies an average of 50.2\% more optimality bugs across all environments.

The contributions of this paper include:
\begin{itemize}[leftmargin=*]
    \item We propose a novel self-referential testing oracle for detecting optimality bugs in DRL agents. 
    \item We develop \toolname, a unified two-stage framework comprising a safety testing stage (which adopts state-of-the-art method CureFuzz) and our novel optimality testing stage, enabling DRL agents to be evaluated for both catastrophic failures and suboptimal decisions in a single pipeline.
    \item We provide extensive experimental evidence demonstrating that \toolname effectively detects a significant number of optimality issues across diverse environments and DRL algorithms. 
\end{itemize}

\section{Preliminaries}
\label{sec:preliminary}

\subsection{Markov Decision Process}

In general, reinforcement learning addresses the problem of learning to control a dynamical system~\cite{sutton1998reinforcement}. Many RL problems are formalized using \textbf{Markov Decision Processes (MDPs)}~\cite{puterman1990markov}, which are defined by a tuple \(\langle S, S_0, A, T, R, \gamma \rangle\) where:
\begin{itemize}[leftmargin=*]
    \item \(S\) is a set of states. A state \(s \in S\) represents a specific situation the agent encounters within the environment.
    \item $S_0 \subseteq S$ is the set of possible starting states. An initial state $s_0 \in S_0$ is selected for each episode.
    \item \(A\) is a set of actions. An action \(a \in A\) is a decision made by the agent that affects the current state.
    \item \(T: S \times A \times S \rightarrow [0, 1]\) is the transition probability function, where \(T(s'|s, a)\) represents the probability of transitioning from state \(s\) to state \(s'\) by taking action \(a\).
    \item \(R: S \times A \rightarrow \mathbb{R}\) is the reward function, where \(R(s,a)\) is the immediate reward received for taking action \(a\) in state \(s\). Typically, the agent's objective is to maximize the cumulative reward over time.
    \item \(\gamma \in (0, 1]\) is a scalar discount factor that determines the present value of future rewards.
\end{itemize}

\noindent\textbf{Markov Property}. The Markov property implies that the future state depends solely on the current state and the action taken without being influenced by the sequence of past events. In this paper, we assume that the environments used in our experiments adhere to the Markov property.

\noindent\textbf{Trajectory}. A trajectory is a sequence of (state, action, reward) tuples.

\noindent\textbf{Episode}. An episode is always a trajectory that represents a complete run from a start state to a terminal state.

\noindent\textbf{Policy}. A policy \(\pi: S \rightarrow A\) is a strategy that specifies the action \(a\) to be taken when in state \(s\). The goal is to discover an optimal policy $\pi^*$ that maximizes the expected cumulative reward over time. This objective can be formalized as the maximization of the expected sum of discounted rewards:
\begin{equation}
    \max_{\pi} \sum_{t=1}^{\infty} \mathbb{E}_{s_t, a_t \sim \pi} \left[ \gamma^t R(s_t, a_t) \right]
\end{equation}
where $R(s_t, a_t)$ represents the reward received at time step $t$ for taking action $a_t$ in state $s_t$, and  $\mathbb{E}$ denotes the expectation over the states and actions under policy $\pi$.

\subsection{Offline Reinforcement Learning}

\begin{figure}[t]
    \centering
    \Description{Illustration of Offline Reinforcement Learning.}
    \includegraphics[width=0.5\columnwidth]{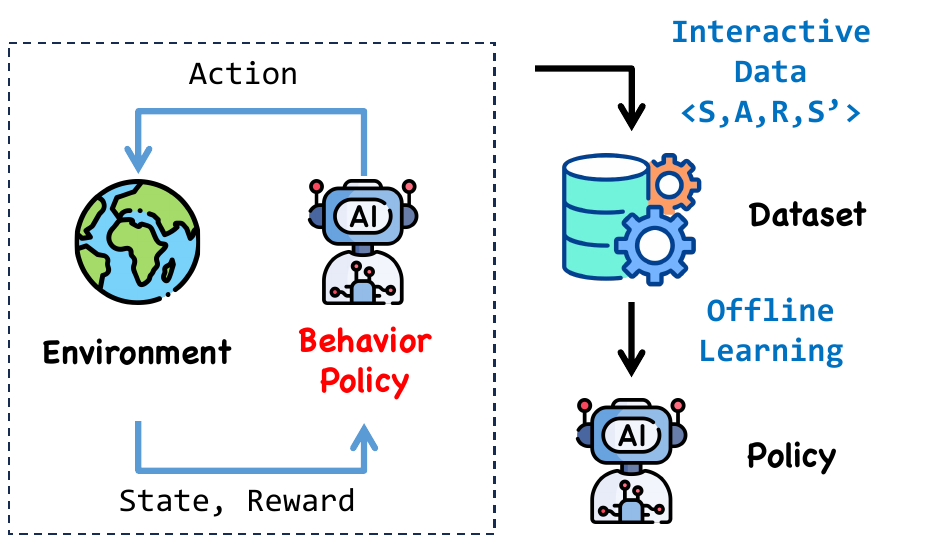}
    \caption{Illustration of Offline Reinforcement Learning.}
    \label{fig:offlinerl}
\end{figure}

Offline Reinforcement Learning (Offline RL)~\cite{levine2020offline, kostrikov2021offline} is an emerging paradigm in RL research. Unlike traditional RL, where agents learn by continuously interacting with the environment, offline RL focuses on creating effective policies exclusively from pre-collected datasets of trajectories without needing further environmental interaction (See Figure~\ref{fig:offlinerl} as an example). This approach is particularly valuable in scenarios where interaction with the environment is expensive, time-consuming, or poses safety risks, such as in healthcare~\cite{tang2021model, tang2022leveraging}, robotics~\cite{sinha2022s4rl, kumar2022pre}, or autonomous driving~\cite{fang2022offline, shi2021offline}. Formally, in the offline RL setting~\cite{levine2020offline, kostrikov2021offline}, learning occurs from a static dataset \(\mathcal{D} = \{(s_i, a_i, r_i, s'_i)\}_{i=1}^N\).  The objective is to learn a policy \(\pi: S \rightarrow A\) using only \(\mathcal{D}\) that maximizes the expected cumulative discounted reward. 
The dataset \(\mathcal{D}\) is generated by an existing policy (or set of policies), which is termed the behavior policy~\cite{sonabend2020expert, fu2020d4rl}. This behavior policy can encompass a wide range of data collection strategies, from random exploration or rule-based approaches to expert demonstrations or policies from previously trained agents.

\section{Approach}
\label{sec:approach}

\begin{figure}[t]
    \centering
    \Description{The Overall Workflow of Delta.}
    \includegraphics[width=0.6\columnwidth]{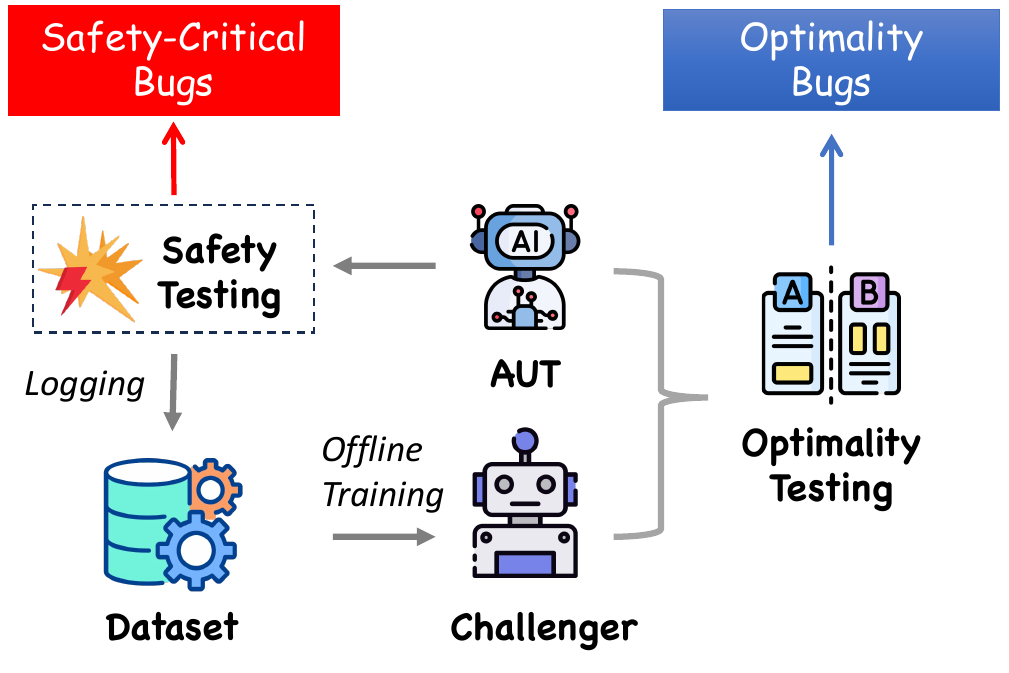}
    \caption{The Overall Workflow of \toolname.}
    \label{fig:workflow}
\end{figure}

\subsection{Assumption}
Our methodology assumes a black-box testing paradigm: we do not require access to the AUT's internal states or parameters. Observations are limited to the AUT's interactions with the environment (i.e., state-action-reward-next\_state tuples) generated during test execution.
Furthermore, we assume the AUT adheres to a deterministic policy and the environment exhibits deterministic transition dynamics.
While the approach is designed for environments modeled as Markov Decision Processes~\cite{puterman1990markov}, it does not necessitate explicit knowledge of the environment's transition dynamics. This characteristic renders the method applicable to scenarios involving opaque AUTs, such as proprietary DRL agents or third-party services.

\subsection{Overview}

\toolname aims to detect both catastrophic failures and optimality bugs in DRL agents. As illustrated in Figure~\ref{fig:workflow}, it consists of two main phases: 

\noindent \textbf{\textit{1) Safety Testing. }} This phase aims to uncover safety-critical issues of the AUT, such as crashes or some hard constraint violations. In our current implementation, we utilize CureFuzz~\cite{he2024curiosity}, a state-of-the-art safety testing method for DRL agents. Notice that \toolname can be seamlessly integrated with any other safety testing frameworks. Concurrently, all AUT behaviors observed during this fuzzing process are collected and stored as a dataset $\mathcal{D}$.

\noindent \textbf{\textit{2) Optimality Testing. }} In the second phase, the dataset $\mathcal{D}$ collected from the safety testing phase is used to train a ``challenger'' agent, utilizing offline RL algorithms. This challenger agent serves as the testing oracle, guiding a differential fuzzing process to detect the AUT's optimality issues. An optimality bug is subsequently flagged in any scenario where the challenger agent achieves a higher cumulative reward than the AUT. We introduce these two phases in detail as follows. 

\subsection{Safety Testing}\label{subsec:safety_testing}
Several safety testing methods have recently been proposed to detect crashes and hard constraint violations in DRL agents~\cite{he2024curiosity, pang2022mdpfuzz, li2023generative}. 
For our safety testing, we did not propose a new safety testing method. In our implementation, we adopt the state-of-the-art testing method, CureFuzz~\cite{he2024curiosity}. 
CureFuzz is a black-box fuzz testing approach. CureFuzz employs a ``curiosity mechanism,'' inspired by Random Network Distillation~\cite{burda2018exploration} and a multi-objective seed selection technique to optimize the fuzzing process to uncover a broad spectrum of crash-triggering scenarios; for details, we refer to the original CureFuzz paper~\cite{he2024curiosity}.

\noindent \textbf{Dataset Construction. }
Concurrently with the safety testing process, all observed trajectories are collected to form a static dataset $\mathcal{D}$. Each trajectory $\tau'$, initiated from a seed $\sigma'$ (via its initial state $s'_0$), consists of a sequence of state-action-reward-next\_state tuples: $\tau' = \{(s_t, a_t, r_t, s_{t+1})\}_{t=0}^{H-1}$, where $H$ is the length of the episode. The dataset $\mathcal{D} = \{\tau'_i\}_{i=1}^N$ thus aggregates all such trajectories recorded during fuzzing; the transition tuples they contain form the static dataset used for offline RL training (Section~\ref{sec:preliminary}).

\subsection{Optimality Testing}

\subsubsection{Overview}
Optimality testing mainly consists of two key stages. First, a \emph{challenger} agent is trained via offline RL algorithms~\cite{levine2020offline, lee2022offline, shi2021offline} on the dataset \(\mathcal{D}\) collected during safety testing. Then, this challenger is employed within a differential fuzzing process to identify optimality bugs of the AUT.

\begin{algorithm}[t]
\small
    \caption{Differential Fuzzing for Optimality Testing}
    \label{alg:differential_fuzzing}
    \begin{algorithmic}
    \STATE \textbf{Input:} Agent Under Test (AUT), Challenger Agent (CA)
    \STATE Initialize seed corpus $\mathcal{S}$; curiosity networks $T$ (fixed), $P$ (trainable)
    \WHILE{time budget not exhausted}
        \STATE Select seed $\sigma$ from $\mathcal{S}$ by energy $E(\sigma)$ (Eq.~\ref{eq:energy_performance}); mutate to $\sigma'$
        \STATE Execute AUT and CA on $\sigma'$; obtain trajectories $\tau'_{\text{aut}}, \tau'_{\text{ca}}$ and rewards $R_{\text{aut}}(\sigma'), R_{\text{ca}}(\sigma')$
        \IF{$R_{\text{ca}}(\sigma') > R_{\text{aut}}(\sigma') + \delta$}
            \STATE Report optimality bug
        \ENDIF
        \STATE Compute diversity $D(\sigma')$ over $\tau'_{\text{aut}}$ (Eq.~\ref{eq:diversity})
        \STATE Update $P$ to minimize $\|T(s) - P(s)\|^2$ over $s \in \tau'_{\text{aut}}$
        \STATE Compute regret $A(\sigma')$ (Eq.~\ref{eq:regret}) and inconsistency $I(\sigma')$ (Eq.~\ref{eq:inconsistency})
        \IF{$D(\sigma') > \bar{D}_{\mathcal{S}}$ \textbf{or} $A(\sigma') > 1$ \textbf{or} $I(\sigma') < \bar{I}_{\mathcal{S}}$}
            \STATE Add $\sigma'$ to $\mathcal{S}$
        \ENDIF
        \STATE Update energy $E(\sigma)$ for all seeds in $\mathcal{S}$
    \ENDWHILE
    \end{algorithmic}
\end{algorithm}

\subsubsection{Oracle Formulation}
\label{subsec:oracle_formulation}

We define an optimality issue as a scenario where the AUT achieves lower cumulative rewards than the challenger agent from the same initial state. The oracle, therefore, relies on the reward function $R$, and its validity depends on what role $R$ plays in our setting.
We treat $R$ as the reference objective for testing---the same objective the AUT was trained against. This follows standard DRL practice~\cite{sutton1998reinforcement}. The design of $R$ is a separate research problem, orthogonal to \toolname.
That said, \toolname is naturally compatible with extended objectives. A practitioner can encode a new non-functional concern into $R$, and rerun \toolname against the extended objective.

Formally, the oracle condition for detecting an optimality issue is:
\begin{equation}
R(\tau_{\text{ca}}) > R(\tau_{\text{aut}}) + \delta
\end{equation}
where $\tau_{\text{aut}}$ and $\tau_{\text{ca}}$ are the trajectories generated by the AUT and the challenger agent from the same initial state $s_0$, $R(\tau)$ denotes the cumulative reward for trajectory $\tau$, and $\delta \geq 0$ is a user-configurable magnitude threshold (default $\delta = 0$, recovering the strict inequality). All experiments in this paper use $\delta = 0$.

\noindent \textbf{Challenger Agent Training.} The dataset $\mathcal{D}$, gathered during the Safety Testing phase (Section~\ref{subsec:safety_testing}), is utilized to train a challenger agent. We employ offline RL algorithms for this training, with specific algorithmic details provided in Section~\ref{sec:experiment}. The quality of the dataset $\mathcal{D}$ is critical to the challenger agent's performance, and the data gathered during safety testing is uniquely suitable for this purpose for several reasons.

First, safety testing naturally explores a wide spectrum of scenarios, including rare edge cases not encountered during standard training. This data diversity is crucial for training a challenger that can generalize across a broad range of situations~\cite{seno2022d3rlpy, suttle2025behavioral}. Second, the dataset contains a rich mixture of learning signals. By including both the AUT's successful trajectories and its documented failures, it provides clear positive and negative examples that guide the challenger on which behaviors to imitate and which to avoid~\cite{levine2020offline, paine2019making}.

Finally, this approach is highly efficient. Since the dataset $\mathcal{D}$ is generated by a competent AUT, it provides a strong baseline that jump-starts the challenger’s training, bypassing the need to learn from scratch. Moreover, the training is conducted entirely offline, requiring no additional, costly environment interactions. Another advantage of \toolname's design is the direct utilization of these existing safety testing logs. This ensures seamless integration with established safety testing and pragmatically eliminates the need for any separate, resource-intensive data collection phase specifically for training the challenger agent.
\begin{figure}[t]
    \centering
    \Description{Delta Differential Fuzzing Workflow.}
    \includegraphics[width=0.8\columnwidth]{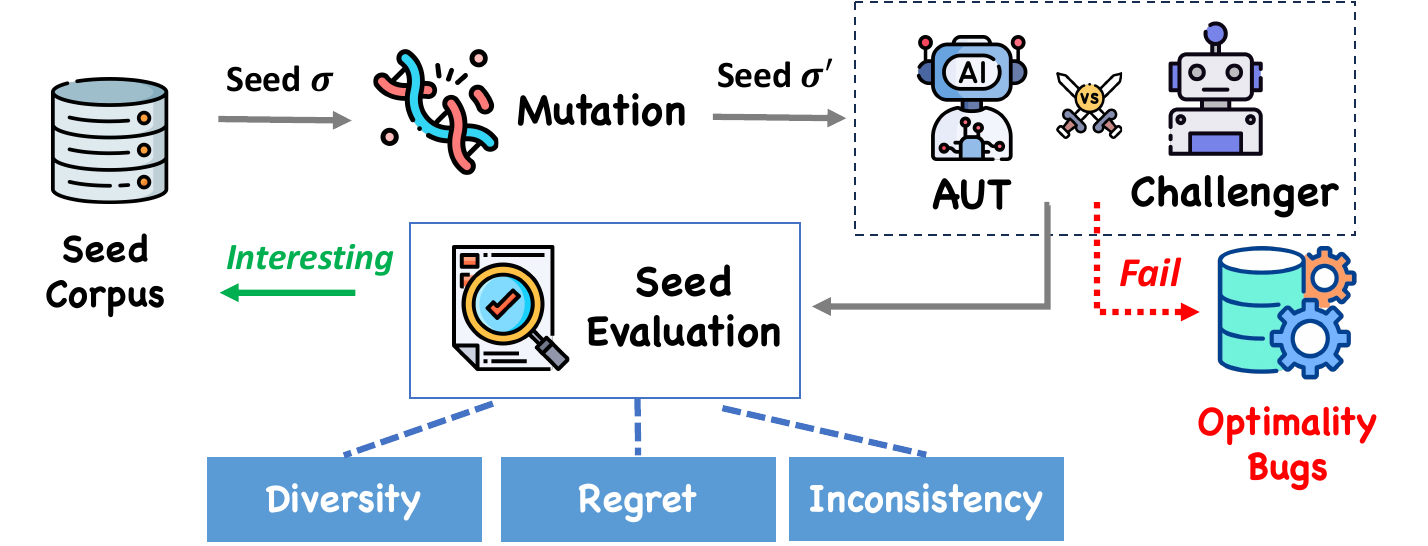}
    \caption{\toolname's Differential Fuzzing Workflow.}
    \label{fig:differential_fuzzing}
\end{figure}

\subsubsection{Differential Fuzzing}

After training the challenger agent, \toolname employs differential fuzzing to identify optimality bugs. As depicted in Figure~\ref{fig:differential_fuzzing}, this process leverages the challenger agent as a testing oracle to systematically evaluate the AUT's decisions.

\noindent \textbf{Overview. } \toolname follows an energy-guided, iterative fuzzing process. To begin, \toolname randomly generates an initial seed corpus, where each seed represents one initial state of the given environment. For example, in autonomous driving, a scenario can be described by the initial position of the autonomous vehicle and surrounding vehicles.
In each iteration, \toolname selects a high energy seed from the corpus and mutates it slightly to generate a new seed (a new initial state). 
\toolname then observes both the AUT and the challenger's actions under the new initial state to assess if the AUT performs suboptimally. 
An optimality bug is flagged if the challenger agent achieves a higher cumulative reward than the AUT. This cycle continues until the allocated time budget is exhausted. To guide the generation of diverse and compelling test cases, \toolname estimates the \emph{energy} of the seed. 
Higher energy seeds, which indicate either novel, unexplored scenarios or greater potential to expose optimality bugs, are prioritized for mutation to produce new test cases. If the mutated seed is deemed ``interesting,'' it would be added back to the seed corpus for consideration in subsequent iterations.

\noindent \textbf{Seed Energy Measurement. }
Seed energy is determined by three factors: \textit{Diversity}, \textit{Regret}, and \textit{Inconsistency}. Hereafter, we denote a seed by $\sigma$. Each seed corresponds to an initial state, typically $s_0$, from which the AUT generates a trajectory. We use $s$ or $s_t$ to represent a generic state within such a trajectory.

 \noindent \textbf{1) Diversity}: Indicating a seed's novelty compared to previously encountered seeds, diversity is key for exploring new AUT behaviors and scenarios. It is quantified using a curiosity mechanism~\cite{he2024curiosity}. The intuition behind this mechanism is to identify states that are ``novel'' to the fuzzer. It employs two neural networks: a \textit{target network} $T$ and a \textit{predictor network} $P$. Both networks share the same architecture (e.g., a Multi-Layer Perceptron (MLP)~\cite{cybenko1989approximation}) and accept states encountered by the AUT as input. The target network $T$ is initialized with random weights and remains fixed throughout the fuzzing process. The predictor network $P$, however, is continuously trained to mimic the output of $T$ using the states observed during fuzzing. 

Note that a randomly initialized DNN is fundamentally different from an unstructured random labeling. Although a DNN's weights may be sampled randomly, the network itself is a deterministic, continuous function of its input: similar inputs produce similar outputs. This locality propagates to the predictor: once $P$ has been trained to mimic $T$ on a set of observed states, it also approximates $T$'s output well for unseen states near them. The prediction error therefore stays low in well-explored regions and rises sharply in unexplored ones, which is precisely what we exploit as a novelty signal. Replacing $T$ with per-state random labels would destroy this property: nearby inputs would carry uncorrelated targets, leaving $P$ with no basis for extrapolation, and every unseen state would yield a similarly large error regardless of its proximity to observed ones.
 
Concretely, for an individual state $s$ encountered in a trajectory, this novelty is measured by the Mean Squared Error~\cite{wackerly2008mathematical} between the outputs of the two networks: $\left\lVert T(s) - P(s) \right\rVert^2$. The diversity of a seed $\sigma$ (with initial state $s_0$), denoted $D(\sigma)$, is then calculated as the average of these novelty scores over all states in the trajectory $\tau$ generated by executing the AUT starting from $s_0$:
\begin{equation}\label{eq:diversity}
   D(\sigma) = \frac{1}{|\tau|} \sum_{s_t \in \tau} \left\lVert T(s_t) - P(s_t) \right\rVert^2
\end{equation}

\noindent \textbf{2) Regret}:
Regret, denoted $A(\sigma)$, measures the performance difference between the AUT and the challenger agent for a specific initial state $s_0$. It quantifies the AUT's suboptimal performance relative to the challenger, reflecting the potential improvement had the AUT followed the challenger's policy. A regret value $A(\sigma) > 1$ indicates that the challenger agent achieves a higher cumulative reward ($R^{\pi_{\text{ca}}}$) than the AUT ($R^{\pi_{\text{aut}}}$) for that scenario. Formally, $A(\sigma)$ is calculated as:
\begin{equation}
\label{eq:regret}
A(\sigma) = \exp \left( R^{\pi_{\text{ca}}}(s_0) - R^{\pi_{\text{aut}}}(s_0) \right)
\end{equation}
This exponential formulation of regret serves to significantly amplify scenarios where the challenger agent markedly outperforms the AUT. By emphasizing this relative performance gain, $A(\sigma)$ guides the fuzzer towards test cases demonstrating clear potential for the AUT's policy improvement, rather than merely towards situations of poor AUT performance that might arise from inherently challenging or unsolvable aspects of the environment for both agents.

\noindent \textbf{3) Inconsistency}: The Inconsistency score, $I(\sigma)$, quantifies the behavioral overlap between the AUT and the challenger agent when starting from the same initial seed $\sigma$; a lower score indicates greater behavioral divergence. It aims to identify scenarios where the two agents explore dissimilar state sequences, potentially revealing instances where the challenger's actions differ significantly from those of the AUT.

To measure inconsistency, we employ Locality Sensitive Hashing (LSH)~\cite{jafari2021survey}. LSH utilizes a hash function $h(s)$ to map high-dimensional states $s$ to lower-dimensional hash codes.
We implement LSH based on Euclidean distance~\cite{danielsson1980euclidean}. The process can be conceptualized as follows:

\begin{enumerate}[leftmargin=*, itemsep=0pt, label=(\alph*)]
    \item \textbf{Random Hyperplane Generation}:
    A set of $k$ random hyperplanes is defined within the high-dimensional state space. Each hyperplane $j$ passes through the origin and is characterized by its random normal vector $r_j$.

    \item \textbf{State Projection and Discretization}:
    For any given state $s$, its projection onto each random normal vector $r_j$ is calculated as the dot product: $p_j(s) = s \cdot r_j$.
    The resulting projected value is then binarized based on its sign. Specifically, a single bit $g_j(s)$ is generated for each hyperplane, indicating on which side of the $j$-th hyperplane the state $s$ lies:
    $$g_j(s) = \begin{cases} 1 & \text{if } p_j(s) > 0 \\ 0 & \text{if } p_j(s) \le 0 \end{cases}$$
    \item \textbf{Binary Hash Code}:
    The resulting sequence of $k$ binary digits, $(g_1(s), g_2(s), \dots, g_k(s))$, forms the discrete hash code, $h(s)$, for state $s$.
\end{enumerate}

The key property is that if two states $s_i$ and $s_j$ are close in Euclidean distance, their projections onto most random lines are also likely to be close, causing them to fall into the same or nearby segments. This results in a high probability that their composite hash codes, $h(s_i)$ and $h(s_j)$, will be identical. Conversely, states far apart in Euclidean distance are more likely to be separated into different segments on many lines, yielding different hash codes.

For a given seed $\sigma$, let $\tau_{\text{aut}}$ and $\tau_{\text{ca}}$ be the trajectories of the AUT and the challenger agent, respectively. We apply the LSH function $h$ to each state in these trajectories, yielding two sets of unique hash codes that represent the distinct regions explored by each agent:
\begin{equation}
H_{\text{aut}} = \{h(s_t) \mid s_t \in \tau_{\text{aut}}\}, 
\quad
H_{\text{ca}} = \{h(s_t) \mid s_t \in \tau_{\text{ca}}\}
\end{equation}

The inconsistency metric $I(\sigma)$ is calculated as the number of LSH hash codes commonly visited by both the AUT and the challenger agent, divided by the total number of unique hash codes visited by the AUT.
\begin{equation}
\label{eq:inconsistency}
I(\sigma) = \frac{|H_{\text{aut}} \cap H_{\text{ca}}|}{|H_{\text{aut}}|}
\end{equation}

This energy function $E(\sigma)$ for optimality testing is defined as:
\begin{equation}
    E(\sigma) = \alpha \cdot D(\sigma) + \beta \cdot A(\sigma) + \lambda \cdot (1 - I(\sigma))
    \label{eq:energy_performance}
\end{equation}
where $D(\sigma)$ is the diversity score, $A(\sigma)$ is the regret (Equation~\ref{eq:regret}), and $I(\sigma)$ is the inconsistency score (Equation~\ref{eq:inconsistency}). The terms $\alpha, \beta, \lambda$ are weighting factors for these components.

\noindent \textbf{Mutation.}
Mutation is the process that generates a new test scenario ($\sigma'$) from an existing seed ($\sigma$) by applying a slight random perturbation to the initial state vector represented by the seed. The nature of this perturbation is adapted to the specific initial state space of each environment.

For instance, in the CartPole environment, a seed's initial state is defined by a vector of four values: cart position, cart velocity, pole angle, and pole angular velocity. To mutate a CartPole seed, a small random value is added to each of these four components, creating a new, slightly different starting condition for the subsequent test run. The specific descriptions of the initial states for all environments are detailed in Section~\ref{sec:experiment}.

\noindent \textbf{Fuzzing.} Algorithm~\ref{alg:differential_fuzzing} provides a detailed description of the differential fuzzing. It mainly consists of two parts: \textit{Initialization} and \textit{Fuzzing Iteration}.

\textit{1) Initialization.}
\toolname generates the initial population of seeds $\mathcal{S}$ by randomly sampling the initial state of the environment based on a uniform distribution. Meanwhile, \toolname initializes the Curiosity module to assign the diversity of each seed in the seed corpus.

\textit{2) Fuzzing Iteration.}
The core fuzzing loop proceeds as follows for each iteration until the time budget is exhausted:
A seed $\sigma$ is selected from the corpus $\mathcal{S}$, where the probability of selecting any seed $\sigma_i$ is proportional to its energy $E(\sigma_i)$ relative to the total energy of all seeds in the corpus:
\begin{equation}
P(\sigma_i) = \frac{E(\sigma_i)}{\sum_{\sigma_j \in \mathcal{S}} E(\sigma_j)} \label{eq:seed_selection_safety}
\end{equation}
The selected seed $\sigma$ is then mutated by applying a slight random perturbation, yielding a new seed $\sigma'$. The AUT and challenger's behavior, when initiated from $\sigma'$ (i.e., its initial state $s'_0$), is subsequently monitored.

\textit{Dynamic Thresholding.}
Manually setting fixed thresholds to identify ``interesting'' seeds for fuzzing can be challenging and often requires domain-specific expertise. Instead, \toolname utilizes a dynamic thresholding mechanism during its differential fuzzing phase to determine which new seeds $\sigma'$ are added to the corpus $\mathcal{S}$. A seed $\sigma'$ is considered interesting and added if it meets at least one of the following criteria:
\begin{enumerate}[leftmargin=*]
    \item Its \textbf{diversity} $D(\sigma')$ exceeds the average diversity $\bar{D}_{\mathcal{S}}$ of seeds currently in the corpus (i.e., $D(\sigma') > \bar{D}_{\mathcal{S}}$), indicating it explores novel states.
    \item Its \textbf{regret} $A(\sigma')$ is greater than 1 (i.e., $A(\sigma') > 1$). This signifies that the challenger agent achieved a higher cumulative reward than the AUT for seed $\sigma'$, directly indicating an instance of AUT suboptimality.
    \item Its \textbf{inconsistency} $I(\sigma')$, which measures trajectory overlap as defined in Equation~\ref{eq:inconsistency}, falls below the average inconsistency $\bar{I}_{\mathcal{S}}$ of seeds in the corpus (i.e., $I(\sigma') < \bar{I}_{\mathcal{S}}$). A lower $I(\sigma')$ value highlights the greater behavioral divergence between the AUT and the challenger agent.
\end{enumerate}
Subsequently, at the end of each iteration, \toolname updates the energy values (calculated using Equation~\ref{eq:energy_performance}) for all seeds remaining in the corpus $\mathcal{S}$. This process of seed selection, mutation, evaluation, and corpus management continues until the allocated time budget is exhausted.

\section{Experiment Setting}
\label{sec:experiment}
\noindent \textbf{Environment}
To assess the feasibility of \toolname, we apply it to multiple case studies, including three classical control environments: CartPole, MountainCar, and Acrobot, as well as two MuJoCo environments: Hopper and Walker2D. These environments are widely used in previous work on testing DRL agents~\cite{zolfagharian2023search, yahmed2023intentional, lu2022towards}.

\begin{itemize}[leftmargin=*]
    \item \textbf{CartPole:} The agent balances a pole on a cart by moving left or right. Success is keeping it upright; episodes last up to 500 steps. Initial states vary by cart position, cart velocity, pole angle, and angular velocity.
    \item \textbf{MountainCar:} The objective is to navigate an underpowered car from a random starting position and zero velocity to a hilltop. The agent controls the car by applying directional forces. The agent receives a reward of -1 per step until success or 200 steps.
    \item \textbf{Acrobot:} The agent swings a two-link pendulum to a target height. It receives -1 reward per step until reaching the target or 500 steps. The initial state involves the cosine and sine values of the angles and their corresponding angular velocities.
    \item \textbf{Hopper:} A one-legged robot aims to hop forward by applying torques to its three actuated joints. Initial states include noise on joint angles and velocities. Rewards are based on forward velocity, survival, and penalties for excessive action. An episode terminates if the Hopper enters an ``unhealthy'' state or after 1000 timesteps.
    \item \textbf{Walker2D:} A bipedal robot learns to walk forward by applying torques to its six actuated joints. Initial states include slight random variations in joint angles and velocities. Similar to Hopper, rewards favor forward progress and upright posture, and episodes end upon entering an ``unhealthy'' state or after 1000 timesteps.
\end{itemize}

\noindent \textbf{Safety Testing Oracle}
A safety violation or catastrophic failure is defined for each environment as follows:
We define safety violations based on standard environment termination criteria~\cite{gymnasium_acrobot, gymnasium_hopper}. For \emph{CartPole}, failure occurs if the pole tilts $>12^\circ$ or the cart moves $>2.4$ units. \emph{MountainCar} fails if the goal is not reached in 200 steps. \emph{Acrobot} is considered unsatisfactory if the reward drops below -100. For \emph{Hopper} and \emph{Walker2D}, a violation occurs if the robot becomes ``unhealthy'' (e.g., falling or exceeding joint limits).

\noindent \textbf{AUT Implementation} We train the AUT using stable and verified implementations of DRL algorithms from the d3rlpy open-source library~\cite{seno2022d3rlpy}. Specifically, for environments with discrete action spaces (CartPole, MountainCar, and Acrobot), we employ the Double Deep Q-Network (DDQN) algorithm~\cite{HasseltGS16}. For environments with continuous action spaces (Hopper and Walker2D), we utilize the Soft Actor-Critic (SAC) algorithm~\cite{haarnoja2018soft}.

We confirmed the high performance of all AUTs prior to safety testing. We report the average cumulative reward and 95\% confidence interval margin of error over 100 runs for each environment, with detailed results presented in Table~\ref{tab:performance}.

\noindent \textbf{Challenger Implementation}
To train the challenger agent, we leverage offline Reinforcement Learning (RL) algorithms, a rapidly evolving subfield of RL. We experiment with three prominent offline RL methods:
\begin{itemize}[leftmargin=*]
    \item \textbf{BC}~\cite{torabi2018behavioral} directly learns a policy by mimicking state-action pairs from a static dataset. 
    \item \textbf{BCQ}~\cite{fujimoto2019off} mitigates distributional shift by constraining the policy to select actions similar to those in the provided data batch.
    \item \textbf{CQL}~\cite{kumar2020conservative} addresses the overestimation of Q-values for out-of-distribution actions by learning a conservative Q-function; it incorporates a regularizer that minimizes Q-values for actions not well-represented in the dataset while maximizing them for dataset actions.
\end{itemize}

For the MuJoCo environments, we utilized hyperparameters from the D4RL replication package~\cite{fu2020d4rl}. Since D4RL does not provide hyperparameters for the classical control environments, we adopted the hyperparameters from the Scope-RL replication package~\cite{kiyohara2023towards}. All experiments were implemented in Python 3.9 and conducted on an Ubuntu 22.04 server equipped with an AMD EPYC 7643 48-core Processor, 504\,GB RAM, and four NVIDIA RTX A5000 GPUs.

\section{Results}
\label{sec:result}

\subsection*{RQ1: Which offline RL algorithm is most effective for training challenger agents to identify optimality bugs?}

\noindent \textbf{Safety Testing.}
We first perform the safety testing phase to identify catastrophic failures in the AUT. Following common practice~\cite{ma2024enhancing, haq2023many, zheng2019wuji}, each run is executed for two hours, which includes a 30-minute sampling phase to construct the initial seed corpus. We perform five independent runs of safety testing for each environment using distinct random seeds. Throughout the entire safety testing process, covering both the sampling and fuzzing phases, every trajectory executed by the AUT is recorded into the dataset $\mathcal{D}$ used for challenger training. Table~\ref{tab:performance} reports, for each environment, the average number of detected failures and the 95\% confidence interval margin.

\noindent \textbf{Performance of Challenger Agents. }
We detail our experimental setup for training challenger agents as follows: 

\textit{1) Dataset Generation:} Safety testing is conducted five times per environment, yielding five distinct datasets for each environment. \textit{2) Challenger Agent Training:}  Each of these five datasets was utilized to train three challenger agents using different offline RL algorithms (BC, BCQ, and CQL). This approach yielded a total of 75 challenger agents (\(5 \text{ environments} \times 5 \text{ datasets} \times 3 \text{ algorithms}\)). \textit{3) Evaluation Metrics:} We calculate the \textit{Mean Cumulative Reward} of one DRL agent under 100 independent runs. The overall performance metrics for a specific offline RL algorithm in a given environment are calculated as the average of its five respective challenger agents. From Table~\ref{tab:performance}, we can observe no single offline RL algorithm uniformly achieves the highest cumulative rewards across all environments; both CQL and BCQ frequently achieved comparable or superior performance. In general, the challenger agents perform very closely to the AUT. This comparable performance implies they are valuable and competent agents for conducting differential analysis with the AUT.

\begin{figure*}[!t]
    \Description{Bar charts of the training time of the AUT and the challenger agents (BC, BCQ, and CQL) in each of the five environments.}
    \centering
    \subfloat[CartPole \label{fig2:cartpole}]{\includegraphics[width = 0.2\linewidth]{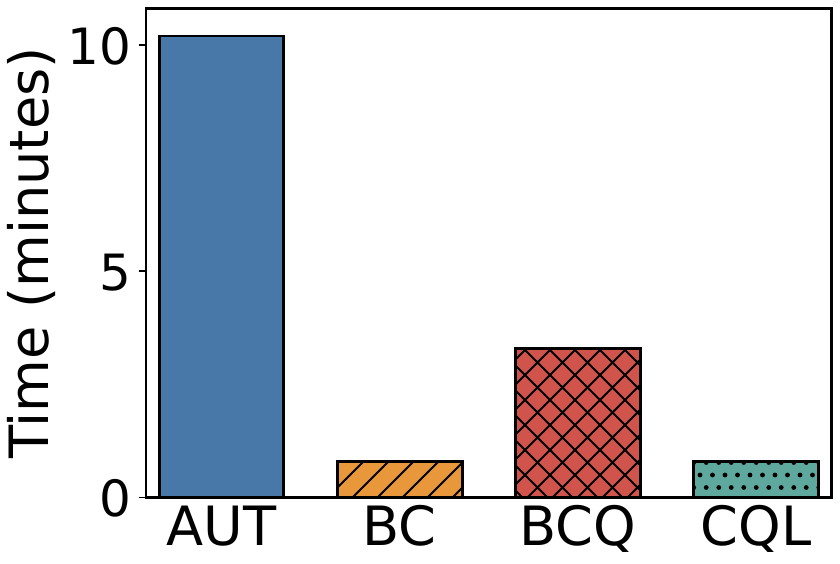}}
    \hfil
    \subfloat[MountainCar \label{fig2:mountaincar}]{\includegraphics[width = 0.2\linewidth]{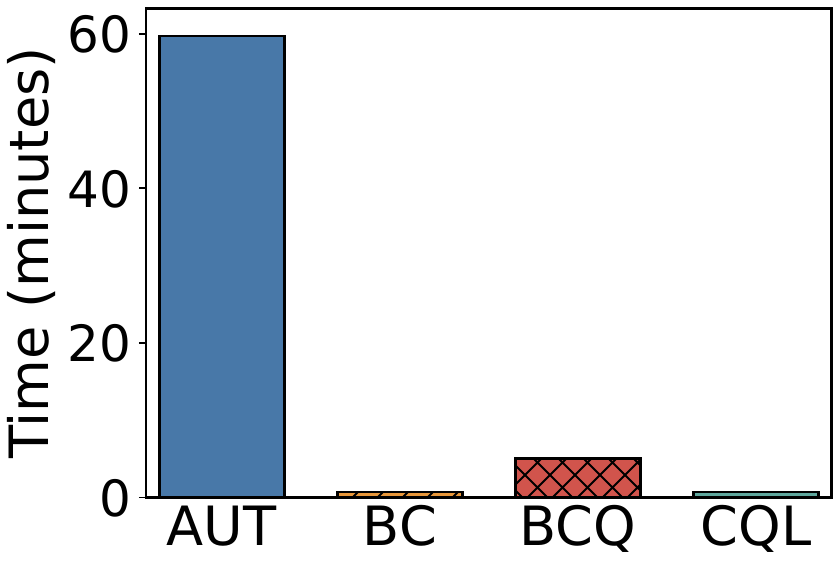}}
    \hfil
    \subfloat[Acrobot \label{fig2:acrobot}]{\includegraphics[width = 0.2\linewidth]{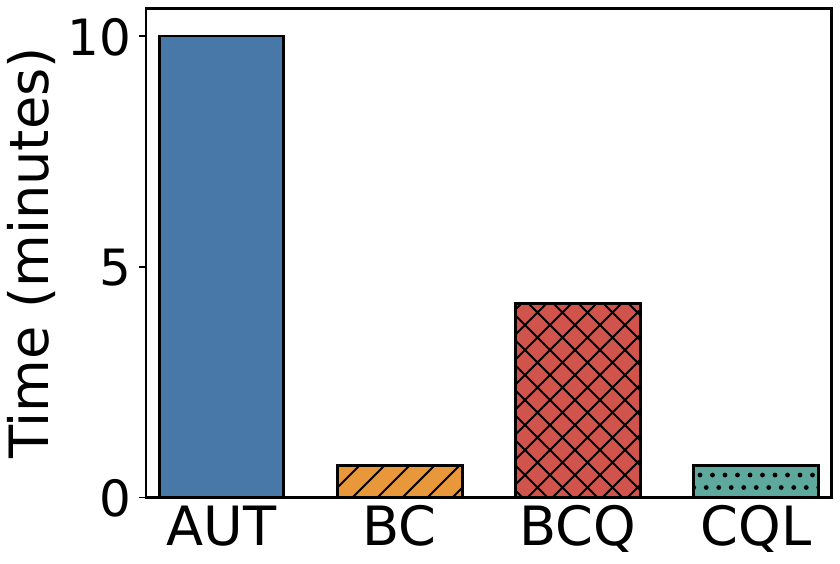}}
    \hfil
    \subfloat[Hopper \label{fig2:hopper}]{\includegraphics[width = 0.2\linewidth]{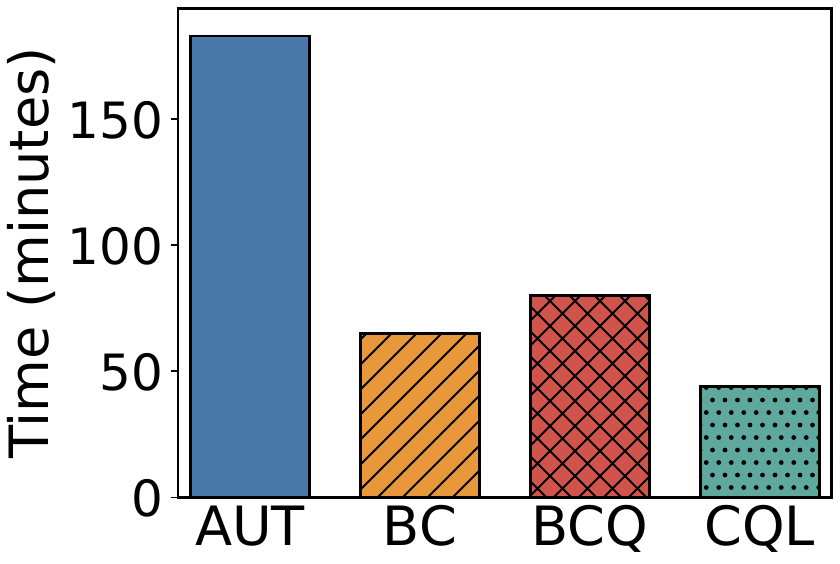}}
    \hfil
    \subfloat[Walker2D \label{fig2:walker2d}]{\includegraphics[width = 0.2\linewidth]{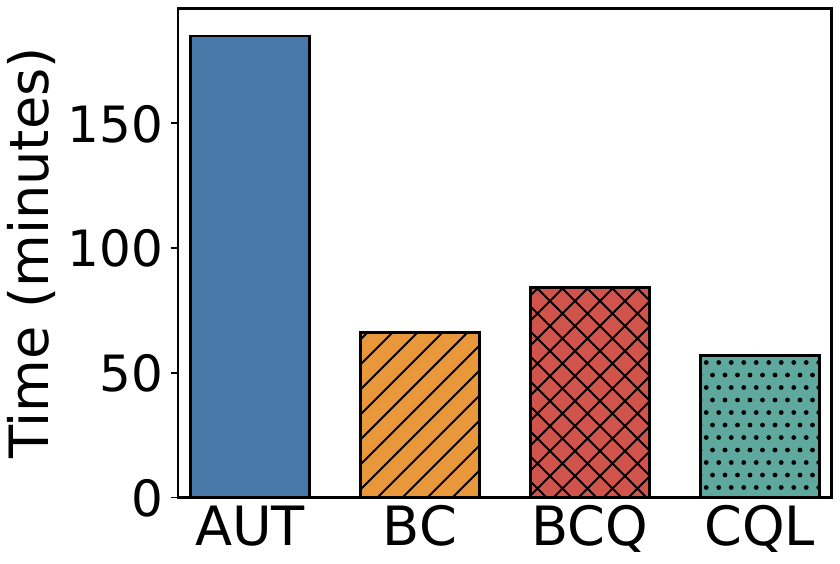}}
    \caption{Running time of training the Agent Under Test (AUT), and the challenger agents (BC, BCQ, and CQL). }
    \label{fig:time}
    \end{figure*}

\noindent \textbf{Efficiency of Training Challenger Agents. } The requirement to train a challenger agent using offline RL may raise concerns about computational overhead. In Figure~\ref{fig:time}, we illustrate the training time of the AUT and challenger agents. Since there are five challenger agents trained for each offline RL algorithm in each environment, we report the median value. However, we notice that the training time is almost identical across runs. In general, the training time for the challenger agents is significantly lower than that of the AUT across all environments. This efficiency gain is mainly due to the different learning paradigms employed~\cite{xie2021policy}. The AUT is trained using traditional reinforcement learning, which requires extensive and often time-consuming interaction with the environment to gather data and learn a policy. In contrast, the challenger agents learn exclusively from a pre-collected, static dataset of trajectories, eliminating the need for any further interaction with the environment.
\begin{table}[tbp]
\centering
\caption{Number of safety failures detected in the AUT, AUT success rate, and mean cumulative reward over 100 episodes for the AUT and challengers (BC, BCQ, CQL). Brackets [lower, upper] denote the 95\% Wilson score CI for the success rate; $\pm$ denotes the 95\% CI margin of error for the other columns. Bold values indicate the best-performing challenger agent (among BC, BCQ, and CQL) for each environment. }
\label{tab:performance}
\resizebox{\columnwidth}{!}{
\begin{tabular}{c|c|c|c|ccc}
\hline
\textbf{Environment} & \textbf{\begin{tabular}[c]{@{}c@{}} Number of Safety\\ Failures of AUT \end{tabular}} & \textbf{\begin{tabular}[c]{@{}c@{}}AUT Success\\ Rate (\%)\end{tabular}} & \textbf{\begin{tabular}[c]{@{}c@{}}AUT\\ Performance\end{tabular}} & \textbf{\begin{tabular}[c]{@{}c@{}}BC\\ Performance\end{tabular}} & \textbf{\begin{tabular}[c]{@{}c@{}}BCQ\\ Performance\end{tabular}} & \textbf{\begin{tabular}[c]{@{}c@{}}CQL\\ Performance\end{tabular}} \\ \hline
\textbf{CartPole}    & $77.4 \pm 10.3$    & $97.0\ [91.6, 99.0]$ & $495.7 \pm 6.8$    & $497.1 \pm 2.5$   & $496.0 \pm 2.6$            & $\textbf{499.8} \pm 1.6$  \\
\textbf{MountainCar} & $167.2 \pm 8.7$    & $99.0\ [94.6, 99.8]$ & $-129.8 \pm 2.8$   & $-129.0 \pm 1.4$  & $-128.2 \pm 1.4$           & $\textbf{-126.6} \pm 1.5$ \\
\textbf{Acrobot}     & $1183.6 \pm 41.9$  & $93.0\ [86.3, 96.6]$ & $-81.3 \pm 3.0$    & $-83.7 \pm 2.1$   & $-84.2 \pm 3.2$            & $\textbf{-83.3} \pm 1.5$  \\
\textbf{Hopper}      & $154.4 \pm 9.9$    & $97.0\ [91.6, 99.0]$ & $3162.3 \pm 60.9$  & $3173.4 \pm 41.6$ & $\textbf{3204.4} \pm 32.4$ & $3173.8 \pm 20.8$         \\
\textbf{Walker2D}    & $251.8 \pm 23.8$   & $95.0\ [88.8, 97.9]$ & $3925.5 \pm 192.2$ & $3961.3 \pm 95.8$ & $\textbf{4004.0} \pm 92.1$ & $3996.5 \pm 68.5$         \\ \hline
\end{tabular}
}
\end{table}

The notably long AUT training time in MountainCar stems from the environment's challenging reward system. In MountainCar, an agent is penalized at every timestep (reward of -1) and only receives a positive reward upon reaching the hilltop goal. This lack of any intermediate positive feedback creates a major exploration challenge: at the beginning of the training, all actions appear equally poor to the agent. It cannot learn incrementally and must instead discover a rare, successful trajectory through prolonged random exploration, which accounts for the significant training time observed.

\noindent \textbf{Optimality Testing.}
Further, we evaluate the effectiveness of the different challenger agents (BC, BCQ, and CQL) in identifying optimality bugs. We conduct five differential fuzzing executions for every combination of environment and offline RL algorithm. In each of these five executions, the AUT is compared against one of the five previously generated challenger agents. Following the same setup as safety testing, each optimality testing execution had a duration of two hours. We focus on the following evaluation metrics:

\begin{itemize}[leftmargin=*]
\item \textbf{\# of Bugs:} The total count of optimality issues.
\item \textbf{\# of Distinct Bugs:} The space of the initial state is discretized into a grid by dividing each dimension into a predefined number of intervals. We use five intervals, following previous work~\cite{ma2024diversity}, except for the MountainCar environment, where 100 intervals are employed due to its simpler nature. The number of Distinct Bugs is then the number of unique grid cells containing at least one detected bug.

\item \textbf{Distance:} Assesses the sparseness of the detected optimality issues within the state space. It is calculated as the average Euclidean distance over all pairs of scenarios where optimality issues were identified.

\end{itemize}

\begin{table}[tbp]
\centering
\caption{Experiment Results of RQ1. We compare the performance of different offline RL algorithms for training the challenger agents to detect optimality bugs. We report the Number of Bugs, the Number of Distinct Bugs, and the Distance. The average result over five runs is reported. The $\pm$ values represent the 95\% confidence interval margin of error. Bold values indicate the best result in each column.}
\label{tab:bugs}
\resizebox{0.9\textwidth}{!}{
\begin{tabular}{c|ccccc}
\hline
\textbf{\begin{tabular}[c]{@{}c@{}}\# of \\ Bugs\end{tabular}}          & \textbf{CartPole}       & \textbf{MountainCar}        & \textbf{Acrobot}                     & \textbf{Hopper}    & \textbf{Walker2D}           \\ \hline
\textbf{BC}                                                             & 79.4 $\pm$ 8.5          & 1243.2 $\pm$ 318.3 & 2900.0 $\pm$ 797.3 & 1231.8 $\pm$ 236.8 & 1251.4 $\pm$ 148.9 \\
\textbf{BCQ}                                                            & \textbf{105.6 $\pm$ 17.7 }       & \textbf{3752.0 $\pm$ 649.9} & \textbf{5450.0 $\pm$ 1241.7}         & \textbf{1738.8 $\pm$ 315.6} & \textbf{1543.2 $\pm$ 146.6} \\
\textbf{CQL}                                                            & 84.4 $\pm$ 10.7         & 2144.0 $\pm$ 797.6 & 4811.0 $\pm$ 760.1          & 875.0 $\pm$ 123.5  & 778.6 $\pm$ 75.7   \\ \hline
\textbf{\begin{tabular}[c]{@{}c@{}}\# of \\ Distinct Bugs\end{tabular}} & \textbf{CartPole}       & \textbf{MountainCar}        & \textbf{Acrobot}            & \textbf{Hopper}    & \textbf{Walker2D}  \\ \hline
\textbf{BC}                                                             & 65.4 $\pm$ 9.8          & 19.8 $\pm$ 6.9     & 298.2$\pm$ 12.2             & 1183.6 $\pm$ 228.8 & 1238.4 $\pm$ 150.8 \\
\textbf{BCQ}                                                            & \textbf{81.0 $\pm$ 9.8   }         & \textbf{58.2 $\pm$ 9.9}     & \textbf{467.4 $\pm$ 10.0   }         & \textbf{1648.8 $\pm$ 280.8} & \textbf{1508.2 $\pm$ 213.4} \\
\textbf{CQL}                                                            & 67.4 $\pm$ 8.5 & 23.8 $\pm$ 5.6              & 432.8$\pm$ 15.4             & 852.8 $\pm$ 115.8  & 771.0 $\pm$ 75.8     \\ \hline
\textbf{Distance}                                                                & \textbf{CartPole}       & \textbf{MountainCar}        & \textbf{Acrobot}            & \textbf{Hopper}    & \textbf{Walker2D}  \\ \hline
\textbf{BC}                                                             & 0.067 $\pm$ 3e-3        & 0.019 $\pm$ 8e-3   & 0.155 $\pm$ 3e-3            & 0.014$\pm$ 2e-4    & 0.018$\pm$ 2e-4    \\
\textbf{BCQ}                                                            & \textbf{0.071$\pm$ 3e-3 }        & \textbf{0.060$\pm$ 2e-2}    & \textbf{0.157$\pm$ 3e-3  }           & \textbf{0.014$\pm$ 2e-4 }   & \textbf{0.019$\pm$ 2e-4 }   \\
\textbf{CQL}                                                            & 0.068$\pm$ 2e-3         & 0.020$\pm$8e-3     & 0.156$\pm$ 3e-3             & 0.014$\pm$ 2e-4    & 0.017$\pm$ 2e-4    \\ \hline
\end{tabular}
}
\end{table}

\noindent \textbf{Results.} For all evaluation metrics, we report the average performance over the five fuzzing executions and the margin of error at 95\% confidence interval. Table~\ref{tab:bugs} demonstrates the capability of BC, BCQ, and CQL in training challenger agents to detect optimality bugs. Across all five environments, the BCQ-trained challenger agents consistently proved to be the most effective at identifying optimality bugs. They detected the highest \emph{Number of Bugs} and \emph{Number of Distinct Bugs} in every environment. For instance, in the MountainCar environment, the BCQ challenger found an average of 3752.0 bugs, significantly outperforming both BC (1243.2) and CQL (2144.0). BCQ-trained challenger agents are therefore established as the default choice to form the oracle when implementing \toolname. 

An interesting insight arises when comparing these bug detection results (Table~\ref{tab:bugs}) with the performance of the challenger agents (Table~\ref{tab:performance}). In cases where CQL-trained challenger agents achieved the highest cumulative rewards, they were less effective at optimality bug detection than BCQ agents. This divergence suggests that the agent characteristics that lead to the highest task rewards are not necessarily the same as those that are most effective for uncovering optimality issues under \toolname.

\begin{tcolorbox}[tile, size=fbox,left=6mm, right=2mm, boxrule=0pt, top=2mm, bottom=2mm,
    borderline west={2mm}{0pt}{blue!70!black}, colback=blue!3!white, 
    sharp corners=south]
    \textbf{Answer to RQ1}: BCQ-trained challenger agents are the most effective for \toolname in detecting optimality bugs, yielding an average of 2,517.9 issues per environment, and are therefore established as the default choice when implementing \toolname.
\end{tcolorbox}

\subsection*{RQ2: How effective is \toolname at detecting optimality bugs compared to baseline methods?}

\noindent \textbf{Baseline Methods.} To the best of our knowledge, \toolname is the first work on this topic and there are no direct baselines. Metamorphic testing (e.g., Decictor~\cite{cheng2024decictor}) relies on manually-defined domain-specific metamorphic relations. General ML testers (e.g., DeepGauge~\cite{ma2018deepgauge}) target static problems. Consequently, we select three state-of-the-art safety testing methods. These methods also prioritize exploration diversity, ensuring broad coverage of the state space.

\begin{itemize}[leftmargin=*]
    \item CureFuzz~\cite{he2024curiosity}: As we described in Section~\ref{sec:approach}, CureFuzz is the state-of-the-art fuzz testing approach to uncover crashes of DRL agents.
    \item MDPFuzz~\cite{pang2022mdpfuzz}: A black-box fuzz testing framework for deep learning models that solve problems modeled as Markov Decision Processes.
    \item GMT~\cite{li2023generative}: A framework that utilizes a generative diffusion model to create test cases and uses novelty-based guidance to diversify agent behaviors.
\end{itemize}

For a fair and direct comparison, all testing methodologies operated under identical conditions: they used the same BCQ-trained challenger agent (from RQ1) to form the differential testing oracle and were allotted a two-hour time budget. To ensure the statistical reliability of our findings, we repeated each experiment five times and report the average performance.

\noindent \textbf{Results.} The experimental results are presented in Table~\ref{tab:baselines}. To summarize, \toolname achieves the best overall performance when compared against the baseline methods across all environments and nearly all evaluation metrics. On average, \toolname identifies 50.2\% more optimality bugs across all environments. Statistical analysis using the Mann-Whitney U test~\cite{mcknight2010mann} and Cohen's d effect size~\cite{cohen1992statistical} confirms that \toolname's superior performance in finding both the total number of bugs and distinct bugs is statistically significant and substantial. Regarding the total Number of Bugs (\# of Bugs), \toolname outperformed all baseline methods in all five environments, consistently identifying the highest total number of optimality issues. This highlights its effectiveness in locating instances of suboptimal agent behavior. The most significant advantage of \toolname is its ability to find a diverse set of unique bugs. It found the highest number of distinct bugs (\# of Distinct Bugs) in every environment, often by a large margin. For example, in the Hopper environment, \toolname found more than double the number of distinct bugs compared to the closest competitor (1648.8 vs. 769.2). In the Acrobot environment, \toolname identifies over five times more distinct bugs than the runner-up (467.4 vs. 93.2). 

On the Distance metric, which measures the spread of the discovered bugs, \toolname's performance is competitive but not universally dominant. It achieved the highest or joint-highest distance score in four of the five environments (MountainCar, Acrobot, Hopper, and Walker2D). The only environment where a baseline method achieved a distinctly better result was CartPole. In CartPole, the generative approach of GMT found fewer distinct bugs than \toolname (43.6 vs. 81.0), yet it achieved the highest distance score (0.084 vs. 0.071). This implies that the few bugs GMT finds are likely to be far apart and unrelated. In contrast, \toolname finds a dense cluster of many unique bugs that are, on average, closer together.

In summary, while not leading in every single metric, the consistent dominance of \toolname in discovering the most total and the most unique bugs makes it the most effective testing framework overall.

\begin{tcolorbox}[tile, size=fbox,left=6mm, right=2mm, boxrule=0pt, top=2mm, bottom=2mm,
    borderline west={2mm}{0pt}{blue!70!black}, colback=blue!3!white, 
    sharp corners=south]
    \textbf{Answer to RQ2}: \toolname significantly outperforms the baseline methods in terms of nearly all evaluation metrics in optimality bug detection. On average, \toolname identifies 50.2\% more optimality bugs.
\end{tcolorbox}

\begin{table}[tbp]
\centering
\caption{Performance comparison of \toolname against baseline methods (MDPFuzz, CureFuzz, and GMT) in detecting optimality bugs, including the total number of bugs found (\# of Bugs), the number of unique bugs (\# of Distinct Bugs), and Distance. All results are averaged over five runs, with the $\pm$ values representing the 95\% confidence interval margin of error. Bold values indicate the best performance in each row. The percentages in parentheses for \toolname denote its improvement over the best-performing baseline for that metric.}
\label{tab:baselines}
\resizebox{0.75\textwidth}{!}{
\begin{tabular}{lcccc}
\hline
\multirow{2}{*}{\textbf{Environment}} & \multicolumn{4}{c}{\textbf{\# of Bugs}}                                                        \\ \cline{2-5} 
                                      & \textbf{\toolname}           & \textbf{MDPFuzz}    & \textbf{CureFuzz}   & \textbf{GMT}        \\ \hline
\textbf{CartPole}                     & \textbf{105.6 $\pm$ 17.7 (+46.7\%)}    & 54.0 $\pm$ 17.6     & 72.0 $\pm$ 8.1      & 60.6 $\pm$ 15.5     \\
\textbf{MountainCar}                  & \textbf{3752.0 $\pm$ 649.9 (+91.5\%)}  & 1854.4 $\pm$ 600.6  & 1959.2 $\pm$ 383.6  & 739.8 $\pm$ 559.1   \\
\textbf{Acrobot}                      & \textbf{5450.0 $\pm$ 1241.7 (+40.9\%) } & 3869.0 $\pm$ 1425.4 & 2556.6 $\pm$ 861.9 & 1850.8 $\pm$ 598.5  \\
\textbf{Hopper}                       & \textbf{1738.8 $\pm$ 315.6 (+37.9\%)}   & 674.2 $\pm$ 101.3   & 815.0 $\pm$160.1    & 1261.2  $\pm$ 286.2 \\
\textbf{Walker2D}                     & \textbf{1543.2 $\pm$ 146.6(+34.1\%)} & 598.6 $\pm$ 141.1   & 838.8 $\pm$ 342.7  & 1150.4  $\pm$ 193.9 \\ \hline
\multirow{2}{*}{\textbf{Environment}} & \multicolumn{4}{c}{\textbf{\# of Distinct Bugs}}                                               \\ \cline{2-5} 
                                      & \textbf{\toolname}           & \textbf{MDPFuzz}    & \textbf{CureFuzz}   & \textbf{GMT}        \\ \hline
\textbf{CartPole}                     & \textbf{81.0 $\pm$ 9.8 (+71.6\%)}        & 43.4 $\pm$ 8.2          & 47.2 $\pm$ 3.9      & 43.6 $\pm$ 7.9     \\
\textbf{MountainCar}                  & \textbf{58.2 $\pm$ 9.9 (+19.3\%)}      & 48.8 $\pm$ 6.7      & 47.8 $\pm$ 10.3     & 35.6 $\pm$ 4.7      \\
\textbf{Acrobot}                      & \textbf{467.4 $\pm$ 10.0 (+401.5\%)}    & 93.2 $\pm$ 14.9     & 75.8 $\pm$ 9.3      & 63.6 $\pm$ 5.2      \\
\textbf{Hopper}                       & \textbf{1648.8 $\pm$ 280.8 (+114.4\%)}  & 641.8 $\pm$ 103.1   & 769.2 $\pm$149.1    & 616.4  $\pm$ 124.7  \\
\textbf{Walker2D}                     & \textbf{1508.2 $\pm$ 213.4 (+81.8\%)}  & 592.4 $\pm$ 140.0   & 829.6 $\pm$ 339.1   & 487.2  $\pm$ 128.7  \\ \hline
\multirow{2}{*}{\textbf{Environment}} & \multicolumn{4}{c}{\textbf{Distance}}                                                          \\  \cline{2-5} 
                                      & \textbf{\toolname}           & \textbf{MDPFuzz}    & \textbf{CureFuzz}   & \textbf{GMT}        \\ \hline
\textbf{CartPole}                     & 0.071 $\pm$ 3e-3    & 0.070 $\pm$ 3e-3    & 0.070 $\pm$ 3e-3    & \textbf{0.084  $\pm$ 3e-3}   \\
\textbf{MountainCar}                  & \textbf{0.060$\pm$ 2e-2}    & 0.052 $\pm$ 1e-2    & \textbf{0.060 $\pm$ 2e-2}    & 0.025 $\pm$ 5e-3    \\
\textbf{Acrobot}                      & \textbf{0.157$\pm$ 3e-3 (+61.9\%)}     & 0.071 $\pm$ 3e-3    & 0.068 $\pm$ 2e-3    & 0.097 $\pm$ 5e-3    \\
\textbf{Hopper}                       & \textbf{0.014$\pm$ 2e-4}    & \textbf{0.014 $\pm$ 1e-4}    & \textbf{0.014 $\pm$ 1e-4}    & 0.010 $\pm$ 8e-4    \\
\textbf{Walker2D}                     & \textbf{0.019$\pm$ 2e-4 (+11.8\%)}     & 0.017 $\pm$ 1e-4    & 0.017 $\pm$ 2e-4    & 0.014 $\pm$ 3e-4   \\
\hline
\end{tabular}
}
\end{table}

\subsection*{RQ3: To what extent does each component of the energy function (i.e., diversity, regret, and inconsistency) contribute to \toolname's optimality bug detection performance?}

To investigate this research question, we implement and experiment with three ablated variants of \toolname (w/o diversity, w/o regret, w/o inconsistency). Table~\ref{tab:ablation} summarizes the experimental results. The removal of the Regret component \textit{(w/o-Regret)} causes the most substantial drop in the total number of bugs found. For example, in the Acrobot environment, the bug count falls by approximately 59\% from 5450.0 to 2238.4. This pattern confirms that prioritizing test cases where the challenger agent shows a clear performance advantage is the single most effective strategy for finding a high volume of optimality bugs. The Diversity component is crucial for discovering a wide range of unique bugs. Its removal \textit{(w/o-Diversity)} leads to a dramatic reduction in the number of Distinct Bugs. This is most clearly seen in the CartPole environment, where the number of distinct bugs is more than halved, dropping from 81.0 to 39.4. Further, removing Inconsistency \textit{(w/o-Inconsistency)} also results in a noticeable decrease in both total and distinct bugs across all environments.

The ablation study concludes that all three energy components, i.e., Regret, Diversity, and Inconsistency, are integral to the success of \toolname. They all play synergistic and distinct roles. Regret is the primary factor for the quantity of bugs.

\begin{table}[tbp]
\centering
\caption{Performance comparison of \toolname against ablated variants (w/o diversity, w/o regret, w/o inconsistency) in detecting optimality bugs, including the total number of bugs found (\# of Bugs), the number of unique bugs (\# of Distinct Bugs), and Distance. All results are averaged over five runs, with the $\pm$ values representing the 95\% confidence interval margin of error. Bold values indicate the best performance in each column.}
\label{tab:ablation}
\resizebox{0.8\textwidth }{!}{
\begin{tabular}{lccccc}
\hline
\multirow{2}{*}{\textbf{Method}} & \multicolumn{5}{c}{\textbf{\# of Bugs}} \\ \cline{2-6}
 & \textbf{CartPole} & \textbf{MountainCar} & \textbf{Acrobot} & \textbf{Hopper} & \textbf{Walker2D} \\ \hline
\textbf{\toolname}        & \textbf{105.6$\pm$17.7} & \textbf{3752.0$\pm$649.9} & \textbf{5450.0$\pm$1241.7} & \textbf{1738.8$\pm$315.6} & \textbf{1543.2$\pm$146.6} \\
w/o-Diversity              & 93.8$\pm$7.7            & 2373.0$\pm$269.4          & 2963.0$\pm$446.4           & 1464.6$\pm$271.6          & 1252.0$\pm$311.7 \\
w/o-Regret                 & 74.6$\pm$11.3           & 2328.0$\pm$252.6          & 2238.4$\pm$165.9           & 1409.8$\pm$214.1          & 1009.4$\pm$247.9 \\
w/o-Inconsistency          & 67.6$\pm$9.0            & 2509.2$\pm$526.3          & 2532.4$\pm$295.2           & 1419.4$\pm$350.9          & 1147.2$\pm$206.0 \\ \hline
\multirow{2}{*}{\textbf{Method}} & \multicolumn{5}{c}{\textbf{\# of Distinct Bugs}} \\ \cline{2-6}
 & \textbf{CartPole} & \textbf{MountainCar} & \textbf{Acrobot} & \textbf{Hopper} & \textbf{Walker2D} \\ \hline
\textbf{\toolname}        & \textbf{81.0$\pm$9.8}   & \textbf{58.2$\pm$9.9}     & \textbf{467.4$\pm$10.0}    & \textbf{1648.8$\pm$280.8} & \textbf{1508.2$\pm$213.4} \\
w/o-Diversity              & 39.4$\pm$2.4            & 43.8$\pm$6.4              & 343.2$\pm$38.8             & 1329.4$\pm$197.9          & 1235.8$\pm$308.7 \\
w/o-Regret                 & 63.0$\pm$8.9            & 48.0$\pm$10.5             & 409.4$\pm$17.2             & 1299.0$\pm$280.8          & 895.6$\pm$240.2 \\
w/o-Inconsistency          & 60.0$\pm$7.7            & 47.6$\pm$10.2             & 448.4$\pm$16.7             & 1315.8$\pm$212.3          & 1037.4$\pm$155.7 \\ \hline
\multirow{2}{*}{\textbf{Method}} & \multicolumn{5}{c}{\textbf{Distance}} \\ \cline{2-6}
 & \textbf{CartPole} & \textbf{MountainCar} & \textbf{Acrobot} & \textbf{Hopper} & \textbf{Walker2D} \\ \hline
\textbf{\toolname}        & \textbf{0.071$\pm$3e-3} & 0.060$\pm$2e-2            & 0.157$\pm$3e-3             & 0.014$\pm$2e-4            & 0.019$\pm$2e-4 \\
w/o-Diversity              & 0.070$\pm$1e-3          & 0.057$\pm$2e-2            & 0.157$\pm$3e-3             & 0.014$\pm$1e-4            & 0.018$\pm$2e-4 \\
w/o-Regret                 & 0.069$\pm$2e-3          & \textbf{0.061$\pm$2e-2}   & \textbf{0.158$\pm$3e-3}    & 0.014$\pm$1e-4            & 0.019$\pm$2e-4 \\
w/o-Inconsistency          & 0.070$\pm$2e-3          & 0.060$\pm$2e-3            & 0.157$\pm$3e-3             & 0.014$\pm$1e-4            & 0.018$\pm$2e-4 \\ \hline
\end{tabular}
}
\end{table}

\begin{tcolorbox}[tile, size=fbox,left=6mm, right=2mm, boxrule=0pt, top=2mm, bottom=2mm,
    borderline west={2mm}{0pt}{blue!70!black}, colback=blue!3!white, 
    sharp corners=south]
    \textbf{Answer to RQ3}: The ablation study demonstrates that Diversity, Regret, and Inconsistency are all vital for \toolname's performance. Regret is the single most crucial factor for maximizing the total number of optimality bugs found.
\end{tcolorbox}

\subsection*{RQ4: How useful are the identified optimality bugs to improve the performance of DRL agents?}

We investigate whether the optimality bugs identified by \toolname can be leveraged to enhance the performance of the AUT. The core strategy involves fine-tuning the AUT using the superior trajectories generated by the challenger agent during the optimality testing phase. The experimental methodology is structured into the following stages:

\noindent (1) \textit{Curation of a Superior Trajectory Dataset}: First, based on the identified optimality bugs, we create a new dataset, $\mathcal{D}_{\text{superior}}$. This dataset is formed by collecting every trajectory where the challenger agent outperformed the AUT. This dataset is thus composed entirely of high-performing trajectories that demonstrate policies superior to the AUT's original behavior.

\noindent (2) \textit{Offline Fine-Tuning}: The original AUTs are further trained using this new dataset. Notice that the AUT cannot be effectively fine-tuned on the static dataset $\mathcal{D}_{\text{superior}}$ using its original online training algorithm (e.g., DDQN or SAC), as these methods perform poorly in offline settings~\cite{lee2022offline, kumar2020conservative, fang2022offline}. Therefore, our approach is to retain the AUT's existing neural network architecture but switch its training objective to one specifically designed for the offline paradigm. We utilize Conservative Q-Learning (CQL) for this purpose. By applying the CQL objective, we can update the agent's existing policy using only the static data, ensuring the fine-tuning process is both stable and robust. Finally, we refer to this fine-tuned agent as \emph{AUT-Improved}.

\noindent \textbf{Result.} We evaluate AUT-Improved in the same environment used to train the original AUT, with no modifications to the transition dynamics, reward function, or action space. The primary metric for comparison is the mean cumulative reward, averaged over 100 independent runs. The experiment results are detailed in Table~\ref{tab:rq4}, where AUT-Improved shows a consistent performance increase against the original AUT across all five environments. The enhancements were particularly notable in several cases: the CartPole agent achieved the theoretical maximum score of 500 with perfect stability ($\pm$~0.0), and the Walker2D agent's performance increased by over 12\%. These results confirm that the optimality bugs identified by \toolname are highly useful for improving cumulative reward. 

\noindent \textbf{Repairing the Safety Failure.} We further experiment with the robustness of AUT-Improved. For each environment, AUT-Improved is evaluated again on the known safety failure-inducing scenarios from the safety testing stage (the safety testing stage from RQ1). The result demonstrates the enhanced robustness of the AUT. AUT-Improved exhibits a substantial reduction in the number of such failures. In CartPole and MountainCar, all safety failures are repaired. Furthermore, AUT-Improved in Walker2D reduces its safety failures by more than 97\%. These results demonstrate that our fine-tuning process not only leads to policies that are higher-performing but also significantly more robust.

\noindent \textbf{Re-evaluation with CureFuzz.} To further validate the robustness of the AUT-Improved, we launch a new fuzzing campaign with CureFuzz against the AUT-Improved. The fuzzing campaign follows the same execution setup described in RQ1. Table~\ref{tab:rq4} presents the results. AUT-Improved consistently exhibits fewer safety failures than the original AUT. Notably, in CartPole, the AUT-Improved is completely robust, with zero safety failures detected. In Acrobot, the number of failures is reduced by approximately 77.6\% (from 1183.6 to 265).

\begin{tcolorbox}[tile, size=fbox,left=6mm, right=2mm, boxrule=0pt, top=2mm, bottom=2mm,
    borderline west={2mm}{0pt}{blue!70!black}, colback=blue!3!white,
    sharp corners=south]
    \textbf{Answer to RQ4}: The identified optimality bugs are highly useful in improving the performance of the AUT. An independent re-evaluation with CureFuzz confirms that the AUT-Improved is substantially more robust, reducing safety failures by up to 100\% across all environments.
\end{tcolorbox}

\begin{table*}[tbp]
\centering
\caption{Comparison of the original Agent Under Test (AUT) and the AUT-Improved. The first two columns report the mean cumulative reward for each agent. The third and fourth columns present the average number of safety failures on known safety-critical scenarios, with percentages in parentheses quantifying the failure reduction after repair. The last column reports the results of the re-evaluation of CureFuzz against the AUT-Improved. All $\pm$ values represent the 95\% confidence interval margin of error.}
\label{tab:rq4}
\resizebox{\textwidth}{!}{
\begin{tabular}{lccccc}
\hline
\textbf{Environment} & \textbf{\begin{tabular}[c]{@{}c@{}}Mean Reward\\ (AUT)\end{tabular}} & \textbf{\begin{tabular}[c]{@{}c@{}}Mean Reward\\ (AUT-Improved)\end{tabular}} & \textbf{\begin{tabular}[c]{@{}c@{}}Safety Failures\\ of AUT\end{tabular}} & \textbf{\begin{tabular}[c]{@{}c@{}}Safety Failures\\ of AUT-Improved\\ (after repair)\end{tabular}} & \textbf{\begin{tabular}[c]{@{}c@{}}CureFuzz Re-eval\\ of AUT-Improved\end{tabular}} \\ \hline
\textbf{CartPole}    & 495.7 $\pm$ 6.8    & 500 $\pm$ 0.0      & 77.4 $\pm$ 10.3   & 0.0(-\textbf{100.0\%}) $\pm$ 0.0   & 0.0 $\pm$ 0.0     \\
\textbf{MountainCar} & -129.8 $\pm$ 2.8   & -125.4 $\pm$ 3.2   & 167.2 $\pm$ 8.7   & 0.0(-\textbf{100.0\%}) $\pm$ 0.0   & 37.4 $\pm$ 6.4  \\
\textbf{Acrobot}     & -81.3 $\pm$ 3.0    & -75.2 $\pm$ 2.9    & 1183.6 $\pm$ 41.9 & 118.2(-\textbf{90.0\%}) $\pm$ 11.3 & 265.0 $\pm$ 18.5  \\
\textbf{Hopper}      & 3162.3 $\pm$ 60.9  & 3247.0 $\pm$ 30.5  & 154.4 $\pm$ 9.9   & 40.0(-\textbf{74.1\%}) $\pm$ 6.0   & 106.2 $\pm$ 19.5 \\
\textbf{Walker2D}    & 3925.5 $\pm$ 192.2 & 4399.3 $\pm$ 93.8  & 251.8 $\pm$ 23.8  & 5.6(-\textbf{97.8\%}) $\pm$ 2.4    & 105.0 $\pm$ 17.2   \\ \hline
\end{tabular}
}
\end{table*}

\section{Discussion}
\label{sec:discussion}

\subsection{Behavior Clustering}
\label{sec:behavior-clustering}

In this subsection, we analyze the AUT's behavior on bug-triggering
initial states. We group the AUT's trajectories from these initial
states into clusters and treat each cluster as a behavioral pattern,
following prior work~\cite{li2024learning}. Specifically, we adopt
unsupervised hierarchical clustering~\cite{murtagh2012algorithms}:
Each trajectory initially forms a singleton cluster, and at every
step, the pair of clusters with the smallest pairwise distance is
merged. This bottom-up process continues until all trajectories
collapse into a single cluster, yielding a dendrogram with maximum
merge distance $d_{\max}$. For pairwise distance, we use Dynamic
Time Warping (DTW)~\cite{berndt1994using} computed over the state
sequences, which natively accommodates trajectories of varying
length without padding. Unlike K-means~\cite{lloyd1982least}, which requires the number of clusters to be fixed in advance, hierarchical clustering imposes
no target count; clusters emerge naturally from the trajectory
structure given a cut threshold $\tau$, expressed as a fraction
of $d_{\max}$. We report results at three values,
$\tau \in \{0.10, 0.20, 0.30\}$, in Table~\ref{tab:nbp_results}.

\begin{table}[t]
\small
\centering
\caption{Number of behavioral patterns detected by \toolname
across the five environments at three relative distance thresholds
$\tau \in \{0.10, 0.20, 0.30\}$. Results are averaged over five
independent runs; $\pm$ values denote the 95\% confidence interval
margin of error.}
\label{tab:nbp_results}
\begin{tabular}{c ccccc}
\toprule
$\boldsymbol{\tau}$ & \textbf{CartPole} & \textbf{MountainCar} & \textbf{Acrobot} & \textbf{Hopper} & \textbf{Walker2D} \\
\midrule
0.10 & 100.2 $\pm$ 15.2 & 4.2 $\pm$ 1.0 & 279.0 $\pm$ 10.5 & 1029.8 $\pm$ 139.0 & 1505.6 $\pm$ 204.6 \\
0.20 & 25.4 $\pm$ 3.4 & 3.2 $\pm$ 0.6 & 130.6 $\pm$ 8.9 & 93.8 $\pm$ 6.4 & 1111.2 $\pm$ 152.7 \\
0.30 & 3.8 $\pm$ 1.4 & 3.0 $\pm$ 0.0 & 54.8 $\pm$ 2.4 & 56.6 $\pm$ 5.9 & 445.2 $\pm$ 116.3 \\
\bottomrule
\end{tabular}
\end{table}

\noindent\textbf{Findings.} The number of behavioral patterns
varies with both the environment and the cut threshold $\tau$,
decreasing monotonically as $\tau$ grows. MountainCar yields only 3–4 behavioral patterns across the three thresholds, reflecting the environment's low-dimensional state space and small discrete action space.
In contrast, higher-dimensional environments
such as Walker2D exhibit hundreds of distinct behavioral patterns
even at $\tau = 0.30$.

\subsection{Extension to Non-Deterministic Environments}
Our experiment setting assumes a deterministic environment, which is the same setting as in previous work~\cite{eniser2022metamorphic, zolfagharian2023search}. However, many 
real-world environments exhibit stochasticity due to random disturbances 
or noisy dynamics. To demonstrate that \toolname remains applicable in 
such settings, we present a natural extension of our oracle formulation 
based on statistical hypothesis testing, and empirically validate it on 
a non-deterministic benchmark.

\noindent \textbf{Statistical Oracle Formulation.} For a given initial 
state $s_0$, we execute both the AUT and the challenger agent $N$ times 
(default $N = 5$), obtaining two sets of cumulative rewards: 
$\mathcal{R}_{\text{aut}} = \{R^{(1)}_{\text{aut}}, \ldots, 
R^{(N)}_{\text{aut}}\}$ and $\mathcal{R}_{\text{ca}} = 
\{R^{(1)}_{\text{ca}}, \ldots, R^{(N)}_{\text{ca}}\}$. An optimality bug 
is reported when the challenger's rewards are statistically significantly 
higher than the AUT's:
\begin{equation}
\label{eq:stochastic_oracle}
\text{OptimalityBug}(s_0) = \mathbb{1}\left[p < \alpha\right]
\end{equation}
where $p$ is the p-value from a one-sided Mann-Whitney U 
test~\cite{mcknight2010mann} with the alternative hypothesis 
$H_1: R_{\text{ca}} > R_{\text{aut}}$, and $\alpha$ is the significance 
level (default $\alpha = 0.05$).

\noindent \textbf{Empirical Validation.} To demonstrate the effectiveness 
of the statistical oracle, we conducted additional experiments on 
\emph{Hopper-Noise}, a non-deterministic variant of Hopper that 
introduces random wind disturbances~\cite{luo2025scalable, rigter2023one}. 
Following the same protocol as RQ1, the AUT achieves a mean reward of 
$3086.9 \pm 61.0$ over 100 episodes, with $477.4 \pm 91.2$ safety failures 
detected during safety testing. For optimality testing, we adopt the same 
two-hour budget per run and repeat each experiment five times, with 
$N = 5$ rollouts per initial state for the statistical oracle. 
Table~\ref{tab:nondet_results} reports the number of optimality bugs 
detected. BCQ remains the strongest offline RL algorithm for training the 
challenger. Using the BCQ challenger, \toolname detects 171\% more 
optimality bugs than the best-performing baseline, confirming that our 
differential testing methodology remains effective and statistically 
sound under non-determinism.

\begin{table}[t]
\centering
\caption{Optimality bug detection on the non-deterministic Hopper-Noise 
environment. Top: comparison of offline RL algorithms for training the 
challenger. Bottom: \toolname against baseline methods. All results are averaged over five runs; $\pm$ values denote 
the 95\% confidence interval margin of error.}
\label{tab:nondet_results}
\resizebox{0.8\textwidth}{!}{
\begin{tabular}{lcccc}
\toprule
\textbf{Challenger Algorithm} & BC & \textbf{BCQ} & CQL \\
\textbf{\# Bugs} & 44.6 $\pm$ 11.3 & \textbf{87.8 $\pm$ 23.3} & 65.2 $\pm$ 22.8 \\
\midrule
\textbf{Testing Method} & \textbf{\toolname} & CureFuzz & GMT & MDPFuzz \\
\textbf{\# Bugs} & \textbf{87.8 $\pm$ 23.3 (+171\%)} & 32.4 $\pm$ 11.3 & 25.2 $\pm$ 6.7 & 24.6 $\pm$ 6.4 \\
\bottomrule
\end{tabular}
}
\end{table}

\subsection{Reliance on Thorough Safety Testing}
\label{sec:reliance-safety-testing}
The challenger's quality depends on the diversity of the safety-testing dataset $\mathcal{D}$ used to train it. As long as safety testing is conducted thoroughly, $\mathcal{D}$ can be trusted to be diverse; if not, this would imply that safety testing itself has failed to adequately explore the AUT's behavior---a far more serious problem than missed optimality issues. \toolname therefore operates under an implicit precondition: it should be applied only when developers are satisfied that safety testing has been rigorous enough to expose the AUT's safety-critical failures. Under this precondition, $\mathcal{D}$ can be reasonably regarded as diverse, and \toolname's optimality testing serves as a complementary layer on top of safety testing. In practice, we adopt a two-hour safety-testing budget per environment, consistent with prior work~\cite{ma2024enhancing, haq2023many, zheng2019wuji}. Developers working with more complex environments may need to extend this budget accordingly. How long a testing campaign should run is itself a long-standing open question in the software testing community~\cite{bohme2018assurances, lipp2023green}.

\subsection{Practical Considerations}

\noindent \textbf{When Can the Challenger Surpass the AUT?}
The challenger can outperform the AUT when: (1) the AUT has not converged to the optimal policy, and room exists for an alternative policy to outperform it in certain regions of the state space; (2) the safety testing dataset naturally contains diverse scenarios including edge cases. This diversity mitigates distributional shift and enables \emph{stitching} of suboptimal trajectory segments into improved policies~\cite{fu2020d4rl}. Overall, the challenger need not outperform the AUT globally. It functions as a \emph{local} oracle: surpassing the AUT from specific initial states suffices to reveal faults.

\noindent \textbf{When the Challenger Fails to Outperform.}
Importantly, this oracle yields value regardless of outcome. A superior challenger exposes optimality issues and guides policy improvement. When the AUT is already near-optimal, the challenger is unlikely to outperform it. This does not indicate oracle failure; rather, it provides empirical confirmation of the high performance of the AUT. The goal of testing is to assess the optimality, not necessarily to find faults. Confirming that the AUT performs well is as valuable as uncovering defects.

\noindent \textbf{Training Efficiency.}
Our efficiency comparison (Section~\ref{sec:result}) reports only 
offline training time, excluding the data collection performed during 
safety testing. Safety testing is a prerequisite for deploying DRL systems, and is not optional, as safety concerns take precedence over optimality. The data collection period is therefore a zero-cost byproduct of an already-required process.

\noindent \textbf{Repair vs. Replace.} RQ4 shows that the optimality bugs detected by \toolname have practical value for repairing the AUT via fine-tuning. A natural question is 
whether the challenger should instead be deployed directly as the final 
agent. We view fine-tuning and replacement as complementary downstream 
choices enabled by \toolname's bug detection. When the AUT's weights are 
inaccessible, fine-tuning is not feasible, and the challenger becomes a candidate for direct deployment. The choice between these modes is ultimately a deployment decision determined by access constraints and operational requirements.

\section{Threats to Validity}
One threat to \emph{internal validity} is the randomness of the testing algorithms used in our experiments. To mitigate it, we repeat each experiment five times and use appropriate statistical tests to account for both statistical significance and effect size. When implementing CureFuzz~\cite{he2024curiosity}, MDPFuzz~\cite{pang2022mdpfuzz}, and GMT~\cite{li2023generative}, we reuse their officially released replication packages. In terms of \emph{external validity}, one threat is that the results of our analysis may not be generalizable. To mitigate it, we have carefully selected diverse tasks and diverse AUTs to evaluate our method, which ensures that our results are not biased. One potential threat to \emph{construct validity} is that the evaluation metrics may not fully capture the performance of our baselines. The number of bugs, distance, and distinct bugs are widely adopted metrics for fuzzing DRL agents~\cite{pang2022mdpfuzz, he2024curiosity, ma2024diversity}.

A further threat to construct validity concerns how optimality
bugs are counted. Following established practice in DRL
testing~\cite{pang2022mdpfuzz, he2024curiosity, ma2024diversity},
we count bugs at the granularity of the initial state, as initial
states are concrete, reproducible inputs that developers can
directly use for debugging and retraining. However, counting at
this granularity does not directly reflect the diversity of
underlying agent behaviors. To complement
this initial-state count, we additionally report a
behavioral-pattern count in
Section~\ref{sec:behavior-clustering}.
Another threat to construct validity concerns the possibility of false negatives. \toolname's challenger is trained on data collected during safety testing; if the safety tester underexplores certain regions of the state space, the challenger may be undertrained in those regions and fail to surpass the AUT even where the AUT is genuinely suboptimal. To mitigate this, \toolname is designed to integrate with any state-of-the-art safety testing method, and our current implementation adopts CureFuzz~\cite{he2024curiosity}, which explicitly promotes exploration diversity via a curiosity-driven mechanism. We further discuss this dependency in Section~\ref{sec:reliance-safety-testing}. Nonetheless, \toolname's reported bug counts should be interpreted as a sound-but-not-complete lower bound on the AUT's optimality issues.

\section{Related Work}
\label{sec:relate}
\noindent\textbf{Safety Testing of DRL Agents.}
Pang et al. introduce MDPFuzz~\cite{pang2022mdpfuzz}, which represents the pioneering effort in black-box fuzz testing for deep learning models addressing Markov Decision Processes, of which DRL agents are a straightforward application. 
He et al. further propose CureFuzz~\cite{he2024curiosity}. CureFuzz proposes a curiosity mechanism to measure the novelty of a scenario, which aims to reveal a diverse set of crash-triggering scenarios. 
Li et al.~\cite{li2023generative} present GMT, a testing framework that utilizes a generative diffusion model~\cite{yang2023diffusion} to generate practical test cases and novelty-based guidance, thereby diversifying agent behaviors and improving test effectiveness. Shi et al.~\cite{shi2025synthify} propose Synthify, which accelerates the falsification of AI-enabled control systems by synthesizing lightweight proxy programs that approximate the neural controller. Complementary to offline testing, they further synthesize efficient and permissive programmatic runtime shields that correct unsafe actions of neural policies at deployment time~\cite{shi2025aegis}. As multi-agent systems are increasingly adopted for complex real-world tasks~\cite{he2025mas}, testing their reliability has attracted growing attention. Ma et al.~\cite{ma2024enhancing} focus on testing multi-agent DRL systems and propose MASTest, which incorporates both individual diversity and team diversity. Furthermore, they propose AdvTest~\cite{ma2024diversity} to evaluate competitive game agents through constraint-guided adversarial agent training.

\noindent\textbf{Search-Based Testing of DRL Agents.}
Zolfagharian et al.~\cite{zolfagharian2023search} propose STARLA, which uses a genetic algorithm to find faulty episodes and actions leading to crashes. 
Tappler et al.~\cite{TapplerCAK22} introduce a search-based testing approach for RL agents with stochastic policies. Their method employs a depth-first backtracking search algorithm to identify reference traces that solve RL tasks and boundary states leading to unsafe states. Tappler et al.~\cite{tappler2024learning} further present a framework that integrates search-based testing with Reinforcement Learning from Demonstrations (RLfD)~\cite{nair2018overcoming}. 

\noindent\textbf{Mutation Testing of DRL Agents.}
Tambon et al.~\cite{tambon2023mutation} and Lu et al.~\cite{lu2022towards} introduce several RL-specific mutation operators. 
Thomas et al.~\cite{thomas2024muprl} introduce muPRL, which utilizes a systematic taxonomy of real-world faults to design specialized mutation operators and evaluate the effectiveness of testing suites in detecting realistic DRL bugs.

\noindent\textbf{Metamorphic Testing of DRL Agents.}
Eniser et al. apply metamorphic testing using manually designed relaxations as metamorphic oracles to assess DRL policies~\cite{eniser2022metamorphic}. Subsequently, Eisenhut et al. introduce new metamorphic oracles alongside a search-based testing method to identify bugs in DRL agents~\cite{eisenhut2023automatic}. More recently, Cheng et al. propose Decictor~\cite{cheng2024decictor}, a method designed to generate non-optimal decision scenarios, specifically in cases where an autonomous driving system fails to plan optimal paths for autonomous vehicles (AVs). A limitation of Decictor is its reliance on pre-existing knowledge of some optimal paths and domain-specific metamorphic relations. In contrast, our work tackles optimality issues using a differential testing approach, making it more broadly applicable.

\section{Conclusion and Future Work}
\label{sec:conclusion}

In this paper, we introduced \toolname, a two-phase testing framework that automatically identifies both safety-critical and optimality bugs. \toolname first conducts safety testing on the AUT while collecting its interaction trajectories. Subsequently, these trajectories are used to train a challenger agent via Offline Reinforcement Learning; if the challenger achieves higher cumulative rewards, an optimality issue is flagged in the AUT. Our extensive evaluations across five diverse environments demonstrated \toolname's effectiveness in uncovering optimality issues. Challenger agents trained with BCQ proved most effective for \toolname in identifying an average of 2,518 optimality issues per environment, and \toolname significantly outperformed baseline methods. In future work, we plan to evaluate \toolname on more complex real-world domains, such as autonomous driving, and to explore emerging offline RL algorithms for training stronger challenger agents.

\section{Data Availability}
\begin{tcolorbox}
A replication package, including source code, experimental datasets, and detailed results, is available on Figshare \cite{he2026delta_artifact} and can be accessed at \url{https://doi.org/10.6084/m9.figshare.29196884.v1}.
\end{tcolorbox}

\balance
\bibliographystyle{ACM-Reference-Format}
\bibliography{reference}

\end{document}